\documentclass[11pt,a4paper]{article}
\usepackage{amsmath}
\usepackage{amssymb}
\usepackage{amsthm}
\usepackage{epsfig}
\usepackage{times}
\usepackage{euler}
\usepackage{wrapfig}
\usepackage{rotating}
\usepackage{a4}
\usepackage[hang,stable]{footmisc}

\newcommand{\sss}{\scriptscriptstyle}

\newcommand{\E}{\mathcal{E}}
\begin{document}
\ifx\href\undefined\else\hypersetup{linktocpage=true}\fi

\newtheorem{theorem}{Theorem}
\newtheorem{proposition}{Proposition}
\newtheorem{Quote}{Einstein Quote}
\theoremstyle{definition}
\newtheorem{Requirement}{Requirement}
\newtheorem{Remark}{Remark}

\title{What is (not) wrong with scalar gravity?}
\author{Domenico Giulini            \\
        Max-Planck-Institute for Gravitational Physics\\
        Albert-Einstein-Institute       \\
        Am M\"uhlenberg 1   \\
        D-14476 Golm, Germany}

\date{}

\maketitle

\begin{abstract}
\noindent
On his way to General Relativity, Einstein gave several arguments 
as to why a special-relativistic theory of gravity based on a 
massless scalar field could be ruled out merely on grounds 
of theoretical considerations. We re-investigate his two main arguments, 
which relate to energy conservation and some form of the principle 
of the universality of free fall. We find that such a theory-based 
\emph{a\,priori} abandonment not to be justified. Rather, the theory seems 
formally perfectly viable, though in clear contradiction with (later)
experiments. 
\end{abstract}

\begin{small}
\setcounter{tocdepth}{3}
\tableofcontents
\end{small}
\newpage
\section{Introduction}
General Relativity (henceforth `GR') differs markedly in many 
structural aspects from all other theories of fundamental 
interactions, which are all formulated as Poincar\'e invariant 
theories in the framework of Special Relativity (henceforth `SR').
The characterisation of this difference has been a central 
theme not only for physicists, but also for philosophers and 
historians of science. Einstein himself emphasised in later (1933) 
recollections the importance of his failure to formulate a viable 
special-relativistic theory of gravity for the understanding of 
the genesis of GR.

Any attempt to give such a characterisation should clearly include 
a precise description of the constraints that prevent gravity from 
also fitting into the framework of SR. In modern terminology, a 
natural way to proceed would be to consider fields according to mass 
and spin\footnote{Mass and spin are the eigenvalues of the so-called 
Casimir operators of the Poincar\'e group, that label its irreducible 
representations.}, discuss their possible equations, the 
inner consistency of the mathematical schemes so obtained, and finally 
their experimental consequences. Since gravity is a classical, 
macroscopically observable, and long-ranged field, one usually 
assumes right at the beginning the spin to be integral and the mass 
parameter to be zero. The first thing to consider would therefore 
be a massless scalar field. What goes wrong with such a theory?

When one investigates this question, anticipating that something does 
indeed go wrong, one should clearly distinguish between the following 
two types of reasonings:
\begin{itemize}
\item[1.]
The theory is internally inconsistent. In a trivial sense this 
may mean that it is mathematically contradictory, in which case 
this is the end of the story. On a more sophisticated level it 
might also mean that the theory violates accepted fundamental 
physical principles, like, e.g., that of energy conservation,
without being plainly mathematically contradictory.    
\item[2.]
The theory is formally consistent and in accord with basic
physical principles. However, it is refuted by experiments.
\end{itemize}
Note that, generically, it does not make much sense to claim both 
shortcomings simultaneously, since `predictions' of inconsistent 
theories should not be trusted. The question to be addressed here 
is whether special-relativistic theories of scalar gravity 
fall under the first category, i.e. whether they can be refuted 
on the basis of formal arguments alone without reference to 
specific experiments. 

Many people think that it can, following A.\,Einstein who accused 
scalar theories to  
\begin{itemize}
\item[a.]
violate some form of the principle of universality of free fall,
\item[b.]
violate energy conservation. 
\end{itemize}
The purpose of this paper is to investigate these statements in detail. 
We will proceed by the standard (Lagrangian) methods of modern 
field theory and take what we perceive as the obvious route when working 
from first principles. 

\section{Historical background}
\label{sec:HistBackground}
As already stressed, the abandonment of scalar theories of gravity by 
Einstein is intimately linked with the birth of GR, 
in particular with his conviction that general covariance must 
replace the principle of relativity as used in SR. 

I will focus on two historical sources in which Einstein complains 
about scalar gravity not being adequate. One is his joint paper 
with Marcel Grossman on the so-called `Entwurf Theory'
(\cite{Einstein:CP}, Vol.\,4, Doc.\,13, henceforth called the 
`Entwurf Paper'), of which Grossmann wrote the ``mathematical part'' 
and Einstein the ``physical part''. Einstein 
finished with \S\,7, whose title asks: ``Can the gravitational 
field be reduced to a scalar\,?''(German original: ``Kann das Gravitationsfeld 
auf einen Skalar zur\"uckgef\"uhrt werden\,?''). In this paragraph 
he presented a Gedankenexperiment-based argument which 
allegedly shows that any special-relativistic scalar theory 
of gravity, in which the gravitational field couples 
exclusively to the matter via the trace of its energy-momentum 
tensor, necessarily violates energy conservation and is hence 
physically inconsistent. This he presented as plausibility argument 
why gravity has to be described by a more complex quantity, like 
the $g_{\mu\nu}$ of the Entwurf Paper, where he and Grossmann 
consider `generally covariant' equations for the first time. 
After having presented his argument, he ends \S\,7 (and his 
contribution) with the following sentences, expressing his 
conviction in the validity of the principle of general 
covariance: 
\begin{Quote}
\label{quote:Einstein1}
Ich mu\ss\ freilich zugeben, da\ss\ f\"ur mich das wirksamste Argument 
daf\"ur, da\ss\ eine derartige Theorie [eine skalare Gravitationstheorie]
zu verwerfen sei, auf der \"Uberzeugung beruht, da\ss\ die Relativit\"at 
nicht nur orthogonalen linearen Substitutionen gegen\"uber besteht, 
sondern einer viel weitere Substitutionsgruppe gegen\"uber. Aber wir 
sind schon desshalb nicht berechtigt, dieses Argument geltend zu machen, 
weil wir nicht imstande waren, die (allgemeinste) Substitutionsgruppe 
ausfindig zu machen, welche zu unseren Gravitationsgleichungen 
geh\"ort.\footnote{%
\textit{To be sure, I have to admit that in my opinion the most effective 
argument for why such a theory [a scalar theory of gravity] has to 
be abandoned rests on the conviction that relativity holds with 
respect to a much wider group of substitutions than just the 
linear-orthogonal ones.  However, we are not justified to push 
this argument since we were not able to determine the (most general) 
group of substitutions which belongs to our gravitational equations.}}
(\cite{Einstein:CP}, Vol.\,4, Doc.\,13, p.\,323)
\end{Quote}

The other source where Einstein reports in more detail on his earlier 
experiences with scalar gravity is his manuscript entitled 
``Einiges \"uber die Entstehung der Allgemeinen Relativit\"atstheorie'', 
dated June\,20th 1933, reprinted in (\cite{Einstein:MeinWeltbild}, pp.\,176-193). 
There he describes in words (no formulae are given) how the `obvious' 
special-relativistic generalisation of the Poisson equation, 
\begin{subequations}
\label{eq:NewtonGravity}
\begin{equation}
\label{eq:NewtonGravity1}
\Delta\Phi=4\pi G\rho\,,
\end{equation}
together with a (slightly less obvious) special-relativistic 
generalisation of the equation of motion,
\begin{equation}
\label{eq:NewtonGravity2}
\frac{d^2\vec x(t)}{dt^2}=-\vec\nabla\Phi(\vec x(t))\,,
\end{equation}
\end{subequations}
lead to a theory in which the vertical acceleration of a test 
particle in a static homogeneous vertical gravitational field 
depends on its initial horizontal velocity and also on 
its internal energy content. In his own words:

\begin{Quote}
\label{quote:Einstein2}
Solche Untersuchungen f\"uhrten aber zu einem Ergebnis, das mich 
in hohem Ma{\ss} mi{\ss}trauisch machte. Gem\"a{\ss} der 
klassischen Mechanik ist n\"amlich die Vertikalbeschleunigung eines
K\"orpers im vertikalen Schwerefeld von der Horizontalkomponente der 
Geschwindigkeit unabh\"angig. Hiermit h\"angt es zusammmen, da{\ss}
die Vertikalbeschleunigung eines mechanischen Systems bzw. dessen 
Schwerpunktes in einem solchen Schwerefeld unabh\"angig herauskommt 
von dessen innerer kinetischer Energie. Nach der von mir versuchten 
Theorie war aber die Unabh\"angigkeit der Fallbeschleunigung von der 
Horizontalgeschwindigkeit bzw. der inneren Energie eines Systems 
nicht vorhanden. Dies pa{\ss}te nicht zu der alten Erfahrung, 
da{\ss} die K\"orper alle dieselbe Beschleunigung in einem 
Gravitationsfeld erfahren. Dieser Satz, der auch als Satz \"uber 
die Gleichheit der tr\"agen und schweren Masse formuliert werden 
kann, leuchtete mir nun in seiner tiefen Bedeutung ein. 
Ich wunderte mich im h\"ochsten Grade \"uber sein Bestehen und 
vermutete, da{\ss} in ihm der Schl\"ussel f\"ur ein tieferes 
Verst\"andnis der Tr\"agheit und Gravitation liegen m\"usse. 
An seiner strengen G\"ultigkeit habe ich auch ohne Kenntnis des 
Resultates der sch\"onen Versuche von E\"otv\"os, die mir -- 
wenn ich mich richtig erinnere -- erst sp\"ater bekannt wurden, 
nicht ernsthaft gezweifelt. Nun verwarf ich den Versuch der oben 
angedeuteten Behandlung des Gravitationsproblems im Rahmer der 
speziellen Relativit\"atstheorie als inad\"aquat. Er wurde offenbar 
gerade der fundamentalsten Eigenschaft der Gravitation nicht 
gerecht. [...] Wichtig war zun\"achst nur die Erkenntnis, da{\ss}
eine vern\"unftige Theorie der Gravitation nur von einer Erweiterung 
des Relativit\"atsprinzips zu erwarten war.\footnote{%
\textit{These investigations, however, led to a result which raised my
strong suspicion. According to classical mechanics, the vertical
acceleration of a body in a vertical gravitational field is
independent of the horizontal component of its velocity.
Hence in such a gravitational field the vertical acceleration
of a mechanical system, or of its centre of gravity, comes out
independently of its internal kinetic energy. But in the theory
I advanced, the acceleration of a falling body was not
independent of its horizontal velocity or the internal energy
of the system. This did not fit with the old experience that 
all bodies experience the same acceleration in a gravitational field. 
This statement, which can be formulated as theorem on the 
equality of inertial and gravitational mass, became clear to me 
in all its deeper meaning. I wondered to the highest degree as to 
why it should hold and conjectured that it be the key for a deeper 
understanding of inertia and gravitation. I did not question its 
rigorous validity, even without knowing about the beautiful 
experiments by E\"otv\"os, of which---if I remember correctly---I 
became aware only later. I now abandoned my attempt as inadequate 
to address the problem of gravitation along the lines outlined above. 
It obviously could not account for the most fundamental property 
of gravitation. [...] The important insight at this stage was that 
a reasonable theory of gravitation could only be expected from 
an extension of the principle of relativity.}} 
(Einstein, 2005, pp.\,178-179)
\end{Quote}

Einstein's belief, that scalar theories of gravity are ruled out, 
placed him---in this respect---in opposition to most of his 
colleagues, like  Nordstr\"om,  Abraham, Mie, and Laue, who took part 
in the search for a (special-) relativistic theory of gravity.
(Concerning Nordstr\"oms theory and the Einstein-Nordstr\"om 
interaction, compare the beautiful discussions by 
Norton~\cite{Norton:1992}\cite{Norton:1993a}.
Some of them were not convinced, it seems, by Einstein's inconsistency 
argument. For example, even after GR was completed, Laue wrote 
a comprehensive review paper on Nordstr\"oms theory, thereby at least 
implicitly claiming inner consistency~\cite{Laue:1917}. Remarkably, 
this paper of Laue's is not contained in his collected writings.

On the other hand, modern commentators seem to be content with a 
discussion of the key r\^ole that Einstein's arguments undoubtedly 
played in the development of GR and, in particular, the requirement 
of general covariance. In fact, already in his famous Vienna 
lecture (\cite{Einstein:CP}, Vol.\,4, Doc.\,17) held on September 23rd 1913, 
less than half a year after the submission of the Entwurf Paper, 
Einstein admits the possibility to sidestep the energy-violation 
argument given in the latter, if one drops the relation between 
space-time distances as given by the Minkowski metric on one hand, 
and physically measured times and lengths on the other. Einstein 
distinguishes between ``coordinate distances'' (German original:
``Koordinatenabstand''), measured by the Minkowski metric, and 
``natural distances'' (German original: ``nat\"urliche Abst\"ande''), 
as measured by rods and clocks (\cite{Einstein:CP}, Vol.\,4, Doc.\,17, p.\,490). 
The relation between these two notions of distance is that of a 
conformal equivalence for the underlying metrics, where the 
``natural'' metric is  obtained from the Minkowski metric by 
multiplying it with a factor that is proportional to the square of 
the scalar gravitational potential. Accordingly, the re-publication 
in January 1914 of the Entwurf Paper includes additional comments, 
the last one of which acknowledges this possibility to sidestep the 
original argument against special-relativistic scalar theories of 
gravity (\cite{Einstein:CP}, Vol.\,4, Doc.\,26, p.\,581). This is sometimes 
interpreted as a ``retraction'' by Einstein of his earlier 
argument (\cite{Einstein:CP}, Vol.\,4, Doc.\,13, p.\, 342, editors 
comment [42]) though Einstein himself speaks more appropriately of 
``evading'' or ``sidestepping'' (German original: ``entgehen''). 
In fact, Einstein does not say that his original argument was erroneous,
but rather points out an escape route that effectively changes the 
hypotheses on which it was based. Indeed, Einstein's re-interpretation of 
space-time distances prevents the Poincar\'e transformations from 
being isometries of space-time, though they formally remain 
symmetries of the field equations. The new interpretation therefore 
pushes the theory outside the realm of SR. Hence Einstein's original 
claim, that a special-relativistic scalar theory of gravity is 
inconsistent, is \emph{not} withdrawn by that re-interpretation. 
Unfortunately, Einstein's recollections do not provide sufficient 
details to point towards a unique theory against which his original 
claim may be tested. But guided by Einstein's remarks and simple first 
principles one can write down a special-relativistic scalar theory
and check whether it really suffers from the shortcomings of the type 
mentioned by Einstein. This we shall do in the main body of this 
paper. We shall find that, as far as its formal consistency is 
concerned, the theory is much better behaved than suggested by 
Einstein. We end by suggesting another rationale (than violation of 
energy conservation), which is also purely intrinsic to the theory 
discussed here, for going beyond Minkowski geometry.

\section{Scalar gravity}
In this section we show how to construct a special-relativistic theory 
for a scalar gravitational field, $\Phi$, coupled to matter. Before we 
will do so in a systematic manner, using variational methods in form 
of a principle of stationary action, we will mention the obvious first 
and naive guesses for a Poincar\'e invariant generalisation of formulae 
(\ref{eq:NewtonGravity}) and point out their deficiencies. 

Our conventions for the Minkowski metric are `mostly minus', that is,  
$\eta_{\mu\nu}=\mbox{diag}(1,-1,-1,-1)$. Given a worldline, $x^\mu(\lambda)$,
where $\lambda$ is some arbitrary parameter, its derivative with 
respect to its eigentime, $\tau$, is written by an overdot,
${\dot x}^\mu:=dx^\mu/d\tau$, where 
$d\tau:=c^{-1}\sqrt{\eta_{\mu\nu}(dx^\mu/d\lambda)(dx^\nu/d\lambda)}\,d\lambda$.
$c$ denotes the velocity of light in vacuum (which we do not set equal 
to unity).  

\subsection{First guesses and a naive theory}
There is an obvious way to generalise the left hand side of 
(\ref{eq:NewtonGravity1}), namely to replace the Laplace operator 
by minus (due to our `mostly minus' convention) the d'Alembert operator:
\begin{equation}
\label{eq:LaplaceToDalembert}
\begin{split}
& \Delta:=
\frac{\partial^2}{\partial x^2}+
\frac{\partial^2}{\partial y^2}+
\frac{\partial^2}{\partial z^2}\\
\mathbf{\mapsto}\quad
-\,&\Box:=
\frac{\partial^2}{\partial x^2}+
\frac{\partial^2}{\partial y^2}+
\frac{\partial^2}{\partial z^2}-
\frac{1}{c^2}\frac{\partial^2}{\partial t^2}
=-\eta^{\mu\nu}\frac{\partial^2}{\partial x^\mu\partial x^\nu}\,.\\
\end{split}
\end{equation}
This is precisely what Einstein reported:

\begin{Quote}
\label{quote:Einstein3}
Das einfachste war nat\"urlich, das Laplacesche skalare Potential 
der Gravitation beizubehalten und die Poisson Gleichung durch ein 
nach der Zeit differenziertes Glied in naheliegender Weise so zu 
erg\"anzen, da{\ss} der speziellen Relativit\"atstheorie Gen\"uge 
geleistet wurde.\footnote{%
\textit{The most simple thing to do was to retain the Laplacian scalar potential 
and to amend the Poisson equation by a term with time derivative, so as to comply 
with special relativity.}}
(Einstein, 2005, p.\,177)
\end{Quote}
Also, the right hand side of (\ref{eq:NewtonGravity1}) need to 
be replaced by a suitable scalar quantity ($\rho$ is not a scalar). 
In SR the energy density is the $00$-component of the energy-momentum 
tensor $T^{\mu\nu}$, which corresponds to a mass density $T^{00}/c^2$. 
Hence a sensible replacement for the right-hand side of 
(\ref{eq:NewtonGravity1}) is:
\begin{equation}
\label{eq:RhoToT}
\rho\quad\mathbf{\mapsto}\quad T/c^2:=\eta^{\mu\nu}T_{\mu\nu}/c^2\,,
\end{equation}
so that (\ref{eq:NewtonGravity1}) translates to 

\begin{equation}
\label{eq:FieldEq}
\Box\Phi=-\kappa T\,,\quad\mathrm{where}\quad
\kappa:=4\pi G/c^2 \,.
\end{equation}
The replacement (\ref{eq:RhoToT}) is not discussed in Einstein's 1933
recollections, but mentioned explicitly as the most natural one
for scalar gravity in Einstein's part of the Entwurf Paper 
(\cite{Einstein:CP}, Vol.\,4, Doc.\,13, p.\,322) and also in his Vienna 
lecture (\cite{Einstein:CP}, Vol.\,4, Doc.\,17, p.\,491). In both cases 
he acknowledges Laue as being the one to draw his attention 
to $T/c^2$ as being a natural choice for the scalar potential's 
source. 

The next step is to generalise (\ref{eq:NewtonGravity2}). 
With respect to this problem Einstein remarks: 
\begin{Quote}
\label{quote:Einstein4}
Auch mu{\ss}te das Bewegungsgesetz des Massenpunktes im 
Gravitationsfeld der speziellen Relativit\"atstheorie angepa{\ss}t 
werden. Der Weg hierf\"ur war weniger eindeutig vorgeschrieben, 
weil ja die tr\"age Masse eines K\"orpers vom Gravitationspotential 
abh\"angen konnte. Dies war sogar wegen des Satzes von der Tr\"agheit 
der Energie zu erwarten.\footnote{%
\textit{Also, the law of motion of a mass point in a gravitational field had to be 
adjusted to special relativity. Here the route was less uniquely mapped out, since the 
inertial mass of a body could depend on the gravitational potential.
Indeed, this had to be expected on grounds of the law of inertia 
of energy.}}
(Einstein, 2005, p.\,177)
\end{Quote}
It should be clear that the structurally obvious 
choice,\footnote{Throughout we write $\nabla_\mu$ for 
$\partial/\partial x^\mu$.}
\begin{equation}
\label{eq:SuperNaiveEqMot}
\ddot x^\mu(\tau)=\eta^{\mu\nu}\nabla_\nu\Phi(x(\tau))\,,
\end{equation}
cannot work. Four-velocities are normed, 
\begin{equation}
\label{eq:FourVelSquare}
\eta_{\mu\nu}\,\dot x^\mu\dot x^\nu=c^2\dot t^2-\dot x^2-\dot y^2-\dot z^2=c^2\,,
\end{equation} 
so that  
\begin{equation}
\label{eq:FourVelFourAcc}
\eta_{\mu\nu}\,\dot x^\mu\ddot x^\nu=0\,.
\end{equation} 
Hence (\ref{eq:SuperNaiveEqMot}) implies the integrability condition 
$\dot x^\mu(\tau)\nabla_\mu\Phi(x(\tau))=d\Phi(x(\tau))/d\tau=0$,
saying that $\Phi$ must stay constant along the worldline of the 
particle, with renders (\ref{eq:SuperNaiveEqMot}) physically totally 
useless. The reason for this failure lies in the fact that we replaced 
the three independent equations (\ref{eq:NewtonGravity2}) by four 
equations. This leads to an over-determination, since the four-velocity 
still represents only three independent functions, due to the 
kinematical constraint (\ref{eq:FourVelSquare}). More specifically, 
it is the component  parallel to the four-velocity $\dot x$ of the 
four-vector equation (\ref{eq:SuperNaiveEqMot}) that leads to the 
unwanted restriction. The obvious way out it to just retain the part 
of (\ref{eq:SuperNaiveEqMot}) perpendicular to $\dot x$:
\begin{subequations}
\label{eq:NaiveEqMot}
\begin{equation}
\label{eq:NaiveEqMot1}
\ddot x^\mu(\tau)
=P^{\mu\nu}(\tau)\nabla_\nu\Phi(x(\tau))\,,
\end{equation}
where $P^{\mu\nu}(\tau)=\eta^{\nu\lambda}P^\mu_{\ \lambda}(\tau)$ 
and  
\begin{equation}
\label{eq:NaiveEqMot2}
P^\mu_\nu(\tau):=
\delta^\mu_\nu-\dot x^\mu(\tau)\dot x_\nu(\tau)/c^2
\end{equation}
\end{subequations}
is the one-parameter family of projectors orthogonal to the four-velocity 
$\dot x(\tau)$, one at each point of the particle's worldline. 
Hence, by construction, this modified equation of motion avoids the 
difficulty just mentioned. We will call the theory based on 
(\ref{eq:FieldEq}) and 
(\ref{eq:NaiveEqMot})
the \emph{naive theory}. We also note that (\ref{eq:NaiveEqMot}) 
is equivalent to 
\begin{equation}
\label{eq:VariableMass1}
\frac{d}{d\tau}\bigl(m(x(\tau))\,\dot x^\mu(\tau)\bigr)
=m(x(\tau))\,\eta^{\mu\nu}\nabla_\nu\Phi(x(\tau))\,,    
\end{equation}
where $m$ is a spacetime dependent mass, given by 
\begin{equation}
\label{eq:VariableMass2}
m=m_0\exp\bigl((\Phi-\Phi_0)/c^2\bigr)\,.
\end{equation}
Here $m_0$ is a constant, corresponding to the value of $m$ 
at gravitational potential $\Phi_0$, e.g., $\Phi_0=0$.  

We could now work out consequences of this theory. However,  
before doing this, we would rather put the reasoning employed 
so far on a more systematic basis as provided by variational 
principles. This also allows us to discuss general matter couplings 
and check whether the matter coupling that the field equation 
(\ref{eq:FieldEq}) expresses is consistent with the coupling 
to the point particle, represented by the equation of motion 
(\ref{eq:NaiveEqMot}). This has to be asked for if we wish to 
implement the equivalence principle in the following form:

\begin{Requirement}[\textbf{Principle of universal coupling}]
All forms of matter (including test particles) couple to 
the gravitational field in a universal fashion.
\end{Requirement} 
\noindent
We will see that in this respect the naive theory is not quite correct. 
We stress the importance of coupling schemes, without which there is 
no logical relation between the field equation and the equation of 
motion for (test-) bodies. This is often not sufficiently taken into 
account in discussions of scalar theories of gravity; 
compare~\cite{Bergmann:1956}\cite{Wellner.Sandri:1964}%
\cite{Harvey:1965a}\cite{Dowker:1965}.

\subsection{A consistent  model-theory for scalar gravity}  
Let us now employ standard variational techniques to establish  
Poincar\'e-invariant equations for the scalar gravitational 
field, $\Phi$, and for the motion of a test particle,so that the 
principle of universal coupling is duly taken care of.    
We start by assuming the field equation (\ref{eq:FieldEq}). 
An action whose Euler-Lagrange equation is (\ref{eq:FieldEq})
is easy to guess\footnote{Note that 
$\Phi$ has the physical dimension of a squared velocity, $\kappa$ 
that of length-over-mass. The pre-factor $1/\kappa c^3$ gives 
the right hand side of (\ref{eq:ActionFieldInt}) the physical 
dimension of an action. The overall signs are chosen according to 
the general scheme for Lagrangians: kinetic minus potential energy.}:
\begin{equation}
\label{eq:ActionFieldInt}
S_{\rm field}+S_{\rm int}= \frac{1}{\kappa c^3}\int d^4x\left(
\tfrac{1}{2}\partial_\mu\Phi\partial^\mu\Phi-\kappa\Phi T\right)\,,
\end{equation}
where $S_{\rm field}$, given by the first term,  is the 
action for the gravitational field and $S_{\rm int}$, 
given by the second term, accounts for the interaction with matter.
 
To this we have to add the action for the matter, $S_{\rm matter}$,
which we only specify insofar as we we assume that the matter consists
of a point particle of rest-mass\footnote{We do not need to indicate 
the rest mass by an additional subscript $0$, since in the sequel we 
never need to distinguish between rest- and dynamical mass. From 
now on $m$ will always refer to rest mass.} $m$ and a `rest' of 
matter that needs not be specified further for our purposes here.  
Hence $S_{\rm matter}=S_{\rm particle}+ S_{\rm rom}$ 
(rom = rest of matter), where
\begin{equation}
\label{eq:ActionPP}
S_{\rm particle}=
-mc^2\int d\tau\,.
\end{equation}
We now invoke the principle of universal coupling to find the 
particle's interaction with the gravitational field. It must be 
of the form $\Phi T_p$, where $T_p$ is the trace of the particle's 
energy momentum tensor. The latter is given by
\begin{equation}
\label{eq:T-PointParticle}
T^{\mu\nu}_p(x)=mc\,\int 
{\dot x}^\mu(\tau){\dot x}^\nu(\tau)\ \delta^{(4)}(x-x(\tau))\ d\tau\,,
\end{equation}
so that the particle's contribution to the interaction term in
(\ref{eq:ActionFieldInt}) is
\begin{equation}
\label{eq:InteractionPP}
S_{\text{int-particle}}=-m\int\Phi(x(\tau))\ d\tau\,.
\end{equation}
Hence the total action can be written in the following form:
\begin{equation}
\label{eq:ScalGrav4}
\begin{split}
S_{\rm tot}=
& - mc^2\int\bigl(1+\Phi(x(\tau))/c^2\bigr)\ d\tau\\
& +\frac{1}{\kappa c^3}\int d^4x\ \bigl(\tfrac{1}{2}\partial_\mu\Phi\partial^\mu\Phi
  -\kappa\Phi T_{\text{rom}}\bigr)\\
& +S_{\text{rom}}\,.
\end{split}
\end{equation}

By construction, the field equation that follows from this action is 
(\ref{eq:FieldEq}), where the energy momentum-tensor refers to the 
matter without the test particle (the self-gravitational field of 
a \emph{test} particle is always neglected). The equations of 
motion for the test particle then turn out to be  
\begin{subequations}
\label{eq:ParticleMotion}
\begin{alignat}{3}
\label{eq:ParticleMotion1}
& &&\ddot x^\mu(\tau)&&\,=\,
P^{\mu\nu}(\tau)\partial_\nu\phi(x(\tau))\,,\\
\label{eq:ParticleMotion2}
&\text{where}\qquad
&&P^{\mu\nu}(\tau)&&\,=\,
\eta^{\mu\nu}-{\dot x}^\mu(\tau){\dot x}^\nu(\tau)/c^2\\
\label{eq:ParticleMotion3}
&\text{and}\qquad
&&\phi&&:\,=\,c^2\ln(1+\Phi/c^2)\,.
\end{alignat}
\end{subequations}
Three things are worth remarking at this point: 
\begin{itemize}
\item
The projector $P^{\mu\nu}$ now appears naturally.
\item
The difference between (\ref{eq:NaiveEqMot}) and 
(\ref{eq:ParticleMotion}) is that in the latter it is $\phi$ 
rather than $\Phi$ that drives the four acceleration. This (only) 
difference to  the naive theory was imposed upon us by the principle 
of universal coupling, which, as we have just seen, determined the 
motion of the test particle. This difference is small for small 
$\Phi/c^2$, since, according to (\ref{eq:ParticleMotion3}), 
$\phi\approx\Phi(1+\Phi/c^2 +\cdots)$. But it becomes essential 
if $\Phi$ gets close to $-c^2$, where $\phi$ diverges and the 
equations of motion become singular. We will see below that 
the existence of the critical value $\Phi=-c^2$ is not necessarily 
a deficiency and that it is, in fact, the naive theory which 
displays an unexpected singular behaviour 
(cf. Section\,\ref{sec:NaiveScalar}).  
\item
The universal coupling of the gravitational field to matter 
only involves the trace of energy-momentum tensor of the latter.
As a consequence of the tracelessness of the pure electromagnetic 
energy-momentum tensor, there is no coupling of gravity to the 
\emph{free} electromagnetic field, like, e.g., a light wave in otherwise 
empty space. A travelling electromagnetic wave will not be influenced 
by gravitational fields. Hence this theory predicts no deflection of 
light-rays that pass the neighbourhoods of stars of other massive 
objects, in disagreement with experimental observations. 
Note however that the interaction of the electromagnetic field with 
other matter will change the trace of the energy-momentum tensor 
of the latter. For example, electromagnetic waves trapped in a 
material box with mirrored walls will induce additional stresses 
in the box's walls due to radiation pressure. This will increase 
the weight of the box corresponding to an additional mass 
$\Delta m=E_{\sss\rm rad}/c^2$, where $E_{\sss\rm rad}$ is the 
energy of the radiation field. In this sense \emph{bound} 
electromagnetic fields \emph{do} carry weight.   
\end{itemize}   

Let us now focus on the equations of motion specialised to static 
situations. That is, we assume that there exists some inertial 
coordinate system $x^\mu$ with respect to which $\Phi$ and hence 
$\phi$ are static, i.e., $\nabla_0\Phi=\nabla_0\phi=0$. We have 
\begin{proposition}
\label{prop:ScalThEqMotNewtonianForm1}
For static potentials (\ref{eq:ParticleMotion}) is equivalent 
to 
\begin{equation}
\label{eq:ScalThEqMotStat1}
\vec x''(t)=-
\bigl(1-\beta^2(t)\bigr)\vec\nabla\phi(\vec x(t))\,,
\end{equation}
where here and below we write a prime for $d/dt$
and use the standard shorthands $\vec v=\vec x'$, 
$\vec\beta=\vec v/c$, $\beta=\Vert\vec\beta\Vert$, and 
$\gamma=1/\sqrt{1-\beta^2}$.
\end{proposition}
\begin{proof}
We write in the usual four-vector component notation: 
$\dot x^\mu=c\gamma(1,\vec\beta)$. Using $d/d\tau=\gamma d/dt$ 
and $d\gamma/dt=\gamma^3(\vec a\cdot\vec v/c^2)$, we have on 
one side
\begin{subequations}
\begin{equation}
\label{eq:ThmStatFieldProof1}
\ddot x^\mu=\gamma^4\,\bigl(\vec a\cdot\vec\beta\,,\,
\vec a_\Vert+\gamma^{-2}\vec a_\perp)\,,
\end{equation}
with $\vec a:=d\vec v/dt$. 
$\vec a_\Vert:=\beta^{-2}\vec\beta(\vec\beta\cdot\vec a)$ and 
$\vec a_\perp:=\vec a-\vec a_\Vert$ are, respectively, the spatial 
projections of $\vec a$ parallel and perpendicular to the 
velocity $\vec v$. On the other hand, we have
\begin{alignat}{1}
\label{eq:ThmStatFieldProof2}
-\dot x^\mu\dot x^\nu\nabla_\nu\phi/c^2
&\,=\,-\gamma^2(\vec\beta\cdot\vec\nabla\phi)(1\,,\,\vec\beta)\,,\\
\label{eq:ThmStatFieldProof3}
\eta^{\mu\nu}\nabla_\nu\phi&\,=\,(0,-\vec\nabla\phi)\,,
\end{alignat}
so that 
\begin{equation}
\label{eq:ThmStatFieldProof4}
\bigl(\eta^{\mu\nu}-\dot x^\mu\dot x^\nu/c^2\bigr)\nabla_\nu\phi
=-\,\gamma^2\bigl(\beta\cdot\nabla\phi\,,\,\vec\nabla_\Vert\phi
+\gamma^{-2}\vec\nabla_\perp\phi\bigr)\,,
\end{equation}
where $\vec\nabla_\Vert:=\beta^{-2}\vec\beta(\vec\beta\cdot\vec\nabla)$
and $\vec\nabla_\perp:=\vec\nabla-\vec\nabla_\Vert$ are the projections 
of the gradient parallel and perpendicular to $\vec v$ respectively.
Equating (\ref{eq:ThmStatFieldProof1}) and (\ref{eq:ThmStatFieldProof4})
results in 
\begin{alignat}{1}
\label{eq:3DimEqMot1}
\vec a\cdot\vec\beta 
&\,=\,-\,\gamma^{-2}\,\vec\beta\cdot\vec\nabla\phi\,,\\
\label{eq:3DimEqMot2}
\vec a &\,=\,-\,\gamma^{-2}\,\vec\nabla\phi\,.
\end{alignat}
Since (\ref{eq:3DimEqMot1}) is trivially implied by 
(\ref{eq:3DimEqMot2}), (\ref{eq:3DimEqMot2}) alone is 
equivalent to (\ref{eq:ParticleMotion}) in the static case, 
as was to be shown.
\end{subequations}
\end{proof}

Einstein's second quote  suggests that he also arrived at an 
equation like (\ref{eq:ScalThEqMotStat1}), which clearly displays 
the dependence of the acceleration in the direction of the 
gravitational field on the transversal velocity. We will come 
back to this in the discussion section.

We can still reformulate (\ref{eq:ScalThEqMotStat1}) so as to look 
perfectly Newtonian (i.e. $m\vec a$ equals a gradient field). 
This will later be convenient for calculating the 
periapsis precession (cf. Sections \ref{sec:PeriPressScalarModel}
and \ref{sec:PeriPressNaiveScalar}).

\begin{proposition}
\label{prop:ScalThEqMotNewtonianForm2}
Let $m$ be the rest-mass of the point particle. 
Then (\ref{eq:ScalThEqMotStat1}) implies  
\begin{equation}
\label{eq:ScalThEqMotStat2}
m\vec a=-\vec\nabla\tilde\phi(\vec x(t))
\quad\mathrm{with}\quad
\tilde\phi:=(mc^2/2)\,\gamma_0^{-2}\,\exp(2\phi/c^2)\,,
\end{equation}
where $\gamma_0$ is an integration constant.
\end{proposition}
\begin{proof}
Scalar multiplication of (\ref{eq:ScalThEqMotStat1}) with 
$\vec v$ leads to 
\begin{equation}
\label{eq:ScalThEqMotStat3}
\bigl(\ln\gamma+\phi/c^2\bigr)'=0\,,
\end{equation}
which integrates to 
\begin{equation}
\label{eq:ScalThEqMotStat4}
\gamma=\gamma_0\,\exp(-\phi/c^2)\,,
\end{equation}
where $\gamma_0$ is a constant. Using this equation to eliminate 
the $\gamma^{-2}$ on the right hand side of (\ref{eq:ScalThEqMotStat1}) 
the latter assumes the form (\ref{eq:ScalThEqMotStat2}).
\end{proof}

\section{Free-fall in static homogeneous fields}
We recall that in Quote\,2 scalar gravity was accused 
of violating a particular form of the principle of the universality 
of free fall, which Einstein called ``the most fundamental property 
of gravitation''. In this section we will investigate the meaning 
and correctness of this claim in some detail. It will be instructive 
to compare the results for the scalar theory with that of a vector 
theory in order to highlight the special behaviour of the former, 
which, in a sense explained below, is just opposite to what Einstein 
accuses it of. We also deal with the naive scalar theory for 
comparison and also to show aspects of its singular behaviour that 
we already mentioned above.

\subsection{The scalar model-theory}
Suppose that with respect to some inertial reference frame with 
coordinates $(ct,x,y,z)$ the gravitational potential $\phi$ just 
depends on $z$. Let at time $t=0$ a body be released at the 
origin, $x=y=z=0$,  with proper velocity $\dot y_0=\dot z_0=0$, 
$\dot x_0=c\beta\gamma$, and $\dot t_0=\gamma$ (so as to obey 
(\ref{eq:FourVelSquare})). As usual $c\beta=v=\dot x_0/\dot t_0$ 
is the ordinary velocity and $\gamma:=1/\sqrt{1-\beta^2}$. We take the 
gravitational field to point into the negative $z$ direction so 
that $\phi$ is a function of $z$ with positive derivative $\phi'$. 
Note that $\dot z(\phi'\circ z)=d(\phi\circ z)/d\tau$ for which 
we simply write $\dot\phi$ with the usual abuse of notion 
(i.e. taking $\phi$ to mean $\phi\circ z$). Finally, we normalise 
$\phi$ such that $\phi(z=0)=0$.
 
The equations of motion (\ref{eq:ParticleMotion1}) now simply 
read 
\begin{subequations}
\label{eq:EqMotVertDrop}
\begin{alignat}{2}
\label{eq:EqMotVertDrop1}
& \ddot t &&\,=\, -\, \dot t\,\dot\phi/c^2\,,\\
\label{eq:EqMotVertDrop2}
& \ddot x &&\,=\, -\, \dot x\,\dot\phi/c^2\,,\\   
\label{eq:EqMotVertDrop3}
& \ddot y &&\,=\, -\, \dot y\,\dot\phi/c^2\,,\\
\label{eq:EqMotVertDrop4}
& \ddot z &&\,=\, -\,\bigl(1+{\dot z}^2/c^2\bigr)\phi'\,.
\end{alignat}
\end{subequations}
The first integrals of the first three equations, keeping 
in mind the initial conditions, are  
\begin{equation}
\label{eq:EqMotDropInt}
\bigl(\dot t(\tau),\dot x(\tau),\dot y(\tau)\bigr)
= \bigl(1,c\beta,0\bigr)\,
\gamma\,\exp\bigl(-\phi(z(\tau))/c^2\bigr)\,.
\end{equation}
Further integration requires the knowledge of $z(\tau)$, that 
is, the horizontal motion couples to the vertical 
one if expressed in proper time.\footnote{In terms of 
coordinate time the horizontal motion decouples: 
$dx/dt=\dot x/\dot t=c\beta \Rightarrow x(t)=c\beta t$.} 
Fortunately, the vertical motion does \emph{not} likewise couple 
to the horizontal one, that is, the right hand side of 
(\ref{eq:EqMotVertDrop4}) just depends on $z(\tau)$. 
Writing it in the form      
\begin{equation}
\label{eq:EqMotVertDrop5}
\frac{\ddot z\dot z/c^2}{1+{\dot z}^2/c^2}=-\dot\phi/c^2
\end{equation}     
immediately allows integration. For $\dot z(\tau=0)=0$ and $\phi(z=h)=0$ 
(so that $\phi(z<h)<0$) we get 
\begin{equation}
\label{eq:EqMotVertDrop6}
\dot z=-c\sqrt{\exp(-2\phi/c^2)-1}\,.
\end{equation}     
From this the eigentime $\tau_h$ for dropping from $z=0$ to $z=-h$ 
with $h>0$ follows by one further integration, showing already at 
this point its independence of the initial horizontal velocity. 

Here we wish to be more explicit and solve the equations of motion 
for the one-parameter family of solutions to (\ref{eq:FieldEq}) for 
$T=0$ and a $\Phi$ that just depends on $z$, namely $\Phi=gz$, for some 
constant $g$ that has the physical dimension of an acceleration. As already 
announced we normalise $\Phi$ such that $\Phi(z=0)=0$. These solutions 
correspond to what one would call a `homogeneous gravitational field'. 
But note that these solutions are \emph{not} globally regular since 
$\phi=c^2\ln(1+\Phi/c^2)=c^2\ln(1+gz/c^2)$ exists only for $z>-c^2/g$ 
and it is the quantity $\phi$ rather than $\Phi$  that 
corresponds to the Newtonian potential (i.e. whose negative gradient 
gives the local acceleration). 

Upon insertion of $\Phi=gz$, (\ref{eq:EqMotVertDrop6}) can be 
integrated to give $z(\tau)$. Likewise, from (\ref{eq:EqMotVertDrop6})
and (\ref{eq:EqMotDropInt}) we can form $dz/dt=\dot z/\dot t$ and 
$dz/dx=\dot z/\dot x$ which integrate to $z(t)$ and $z(x)$ 
respectively. The results are 
\begin{subequations}
\label{eq:ScalSolEqMot}
\begin{alignat}{2}
\label{eq:ScalSolEqMot1}
& z(\tau) &&\,=\,-\,
\frac{c^2}{g}\left\{1-\sqrt{1-\bigl(\tau g/c\bigr)^2}\right\}\,,\\
\label{eq:ScalSolEqMot2}
& z(t) &&\,=\,-\,
\frac{2c^2}{g}\sin^2\bigl(gt/2\gamma c\bigr)\,,\\
\label{eq:ScalSolEqMot3}
& z(x) &&\,=\,-\,
\frac{2c^2}{g}\sin^2\bigl(gx/2\beta\gamma c^2\bigr)\,.
\end{alignat}
\end{subequations}
For completeness we mention that direct integration of 
(\ref{eq:EqMotDropInt}) gives for the other component 
functions, taking into account the initial conditions 
$t(0)=x(0)=y(0)=0$: 
\begin{equation}
\label{eq:EqMotVertDrop9}
\bigl(t(\tau), x(\tau), y(\tau)\bigr)
= \bigl(1,c\beta,0\bigr)\,
(\gamma c/g)\,\sin^{-1}\bigl(g\tau/c\bigr)\,.
\end{equation}
The relation between $\tau$ and $t$ is 
\begin{equation}
\label{eq:EqMotVertDrop10}
\tau\,=\,(c/g)\sin\bigl(gt/\gamma c\bigr)\,.
\end{equation}
Inversion of (\ref{eq:ScalSolEqMot1}) and (\ref{eq:ScalSolEqMot2})
leads, respectively, to the proper time, $\tau_h$, and coordinate time, $t_h$,  
that it takes the body to drop from $z=0$ to $z=-h$:
\begin{subequations}
\label{eq:ScalDropTimes} 
\begin{alignat}{4}
\label{eq:ScalDropEigenTime}
& \tau_h &&\,=\,
\frac{c}{g}\sqrt{1-\bigl(1-gh/c^2\bigr)^2}
&&\,\approx\,&&\sqrt{2h/g}\,,\\
\label{eq:ScalDropCoordTime}
& t_h    &&\,=\,
\gamma\,\frac{2c}{g}\,\sin^{-1}\Bigl(\sqrt{gh/2c^2}\Bigr)
&&\,\approx\,\gamma\,&&\sqrt{2h/g}\,.
\end{alignat}
\end{subequations}
The approximations indicated by $\approx$ refer to the leading order 
contributions for small values of $gh/c^2$ (and any value of 
$\gamma$). The appearance of $\gamma$ in (\ref{eq:ScalDropCoordTime}) 
signifies the quadratic dependence on the initial horizontal 
velocity: the greater the inertial horizontal velocity, the longer 
the span in inertial time for dropping from $z=0$ to $z=-h$. 
This seems to be Einstein's point (cf. Quote\,\ref{quote:Einstein2}). 
In contrast, there is no such dependence in (\ref{eq:ScalDropEigenTime}), 
showing the independence of the span in \emph{eigen}time from the 
initial horizontal velocity.

The eigentime for dropping into the singularity at $z=-h=-c^2/g$ is 
$\tau_{*}=c/g$. In particular, it is  finite, so that a freely 
falling observer experiences the singularity of the gravitational 
field $-\vec\nabla\phi$ in finite proper time. We note that this 
singularity is also present in the static spherically symmetric 
vacuum solution $\Phi(r)=-Gm/r$ to (\ref{eq:FieldEq}), for which 
$\phi(r)=c^2\ln(1+\Phi/c^2)$ exists only for $\Phi>c^2$, i.e. 
$r>Gm/c^2$. The Newtonian acceleration diverges as $r$ approaches 
this value from above, which means that stars of radius smaller 
than that critical value cannot exist because no internal pressure 
can support the infinite inward pointing gravitational pull.
Knowing GR, this type of behaviour does not seem 
too surprising after all. Note that we are here dealing with a 
non-liner theory, since the field equations (\ref{eq:FieldEq})
become non-liner if expressed in terms of $\phi$ according to 
(\ref{eq:ParticleMotion3}).

\subsection{The naive scalar theory}
\label{sec:NaiveScalar}
Let us for the moment return to the naive theory, given by 
(\ref{eq:FieldEq}) and (\ref{eq:NaiveEqMot}). Its equations
of motion in a static and homogeneous vertical field are obtained 
from (\ref{eq:EqMotVertDrop}) by setting $\phi=gz$. Insertion 
into (\ref{eq:EqMotVertDrop6}) leads to $z(\tau)$.  The expressions 
$z(t)$ and $z(x)$ are best determined directly by integrating 
$dz/dt=\dot z/\dot t$ using (\ref{eq:EqMotVertDrop6}) and 
(\ref{eq:EqMotVertDrop9}). One obtains
\begin{subequations}
\label{eq:NaiveScalSolEqMot}
\begin{alignat}{3}
\label{eq:NaiveScalSolEqMot1}
& z(\tau) &&\,=\,-\,
&&\frac{c^2}{g}\,\ln\Bigl(\cos\bigl(g\tau/c\bigr)\Bigr)\,,\\
\label{eq:NaiveScalSolEqMot2}
& z(t) &&\,=\,-\,
&&\frac{c^2}{g}\,\ln\Bigl(\cosh\bigl(gt/\gamma c\bigr)\Bigr)\,,\\
\label{eq:NaiveScalSolEqMot3}
& z(x) &&\,=\,-\,
&&\frac{c^2}{g}\,\ln\Bigl(\cosh\bigl(gx/\beta\gamma c^2\bigr)\Bigr)\,.
\end{alignat}
\end{subequations}
The proper time and coordinate time for dropping from $z=0$ to $z=-h$ 
are therefore given by    
\begin{subequations}
\label{eq:NaiveScalDropTimes}
\begin{alignat}{3}
\label{eq:NaiveScalDropEigenTime}
& \tau_h &&\,=\,\frac{c}{g}\,\cos^{-1}\Bigl(\exp\bigl(-hg/c^2\bigr)\Bigr)
&&\,\approx\,\sqrt{2h/g}\,,\\
\label{eq:NaiveScalDropCoordTime}
& t_h &&\,=\,
\frac{c}{g}\,\gamma\,\cosh^{-1}\Bigl(\exp\bigl(hg/c^2\bigr)\Bigr)
&&\,\approx\,\gamma\,\sqrt{2h/g}\,,
\end{alignat}
\end{subequations}
where $\approx$ gives again the leading order contributions for 
small $gh/c^2$.\footnote{To see this use  
the identity $\cos^{-1}(x)=\tan^{-1}\bigl(\sqrt{x^{-2}-1}\bigr)$.}
The general relation between $\tau$ and $t$ is obtained by inserting 
(\ref{eq:NaiveScalSolEqMot1}) into the expression 
(\ref{eq:EqMotDropInt}) for $\dot t$ and integration:  
\begin{equation}
\label{eq:NaiveScalRelTauT}
\tau=\frac{2c}{g}\,
\left\{\tan^{-1}\bigl(\exp(gt/\gamma c)\bigr)-\pi/4\right\}\,.
\end{equation}

Note that (\ref{eq:NaiveScalDropEigenTime}) is again independent 
of the initial horizontal velocity, whereas 
(\ref{eq:NaiveScalDropCoordTime}) again is not. Moreover, the 
really surprising feature of (\ref{eq:NaiveScalDropEigenTime}) is 
that $\tau_h$ stays finite for $h\rightarrow\infty$. In fact, 
$\tau_\infty=c\pi/2g$. So even though the solution $\phi(z)$ is 
globally regular, the solution to the equations of motion is in a 
certain sense not, since the freely falling particle reaches the 
`end of spacetime' in finite proper time. This is akin to `timelike 
geodesic incompleteness', which indicates singular space-times in GR. 
Note that it need not be associated with a singularity of the 
gravitational field itself, except perhaps for the fact that the 
very notion of an infinitely extended homogeneous field is itself 
regarded as unphysical.

\subsection{Vector theory }
For comparison it is instructive to look at the corresponding 
problem in a vector (spin\,1) theory, which we here do not wish to 
discuss in detail. It is essentially given by Maxwell's 
equations with appropriate sign changes to account for the 
attractivity of like `charges' (here masses). This causes 
problems, like that of runaway solutions, due to the possibility to 
radiate away negative energy. But the problem of free fall in a 
homogeneous gravitoelectric field can be addressed, which is formally 
identical to that of free fall of a charge $e$ and mass $m$ 
in a static and homogeneous electric field $\vec E=-E\vec e_z$. 
So let us first look at the electrodynamical problem. 

The equations of motion (the Lorentz force law) are
\begin{equation}
\label{eq:LorentzEqMotion}
m{\ddot z}^\mu=e\eta^{\mu\nu}F_{\nu\lambda}{\dot z}^\lambda\,,
\end{equation}
where $F_{03}=-F_{30}=-E/c$ and all other components vanish.
Hence, writing 
\begin{equation}
\label{eq:DefCalE}
\E:=eE/mc\,,
\end{equation}
we have  
\begin{subequations}
\label{eq:VectEqMot}
\begin{alignat}{2}
\label{eq:VectEqMot1}
&c\ddot t &&\,=\,-\,\E\dot z\,,\\
\label{eq:VectEqMot2}
&\ddot x &&\,=\,0\,,\\
\label{eq:VectEqMot3}
&\ddot y &&\,=\,0\,,\\
\label{eq:VectEqMot4}
&\ddot z &&\,=\,-\,\E\,c\dot t\,.
\end{alignat}
\end{subequations}
With the same initial conditions as in the scalar case we immediately have 
\begin{equation} 
\label{eq:VectMotSol1}
x(\tau)=c\beta\gamma\tau\,,\quad y(\tau)=0\,.
\end{equation}
(\ref{eq:VectEqMot1}) and (\ref{eq:VectEqMot4}) are equivalent to 
\begin{equation} 
\label{eq:VectMotSol2a}
(c\ddot t\pm \ddot z)=\mp\E(c\dot t\pm \dot z)\,,
\end{equation}
which twice integrated lead to  
\begin{equation} 
\label{eq:VectMotSol2b}
ct(\tau)\pm z(\tau)=A_{\pm}\exp(\mp\E\tau)+B_\pm\,,
\end{equation}
where $A_+,A_-,B_+$, and $B_-$ are four constants of integration. 
They are determined by $z(0)=\dot z(0)=t(\tau)=0$ and 
$c\dot t^2-\dot x^2-\dot y^2-\dot z^2=c^2$, leading to  
\begin{equation}
\label{eq:VectMotRelTauT}
t(\tau)=(\gamma/\E)\,\sinh(\E\tau)
\end{equation}
and also 
\begin{subequations}
\label{eq:VectSolEqMot}
\begin{equation}
\label{eq:VectSolEqMot1}
z(\tau)=-\,\,(2c\gamma/\E)\,\sinh^2(\E\tau/2)\,.
\end{equation}
Using (\ref{eq:VectMotRelTauT}) and (\ref{eq:VectMotSol1}) to eliminate 
$\tau$ in favour of $t$ or $x$ respectively in (\ref{eq:VectSolEqMot1}) 
gives  
\begin{alignat}{2}
\label{eq:VectSolEqMot2}
& z(t) &&\,=\,-\,
\frac{\gamma c}{\E}\left(\sqrt{1+(t\E/\gamma)^2}-1\right)\,,\\
\label{eq:VectSolEqMot3}
& z(x) &&\,=\,-\,
\frac{2\gamma c}{\E}\,\sinh^2(x\E/2\beta\gamma c)\,.
\end{alignat}
\end{subequations}
Inverting (\ref{eq:VectSolEqMot1}) and (\ref{eq:VectSolEqMot2}) gives 
the expressions for the spans of eigentime and inertial time, 
respectively, that it takes for the body to drop from $z=0$ to $z=-h$:
\begin{subequations}
\label{eq:VectDropTimes}
\begin{alignat}{3}
\label{eq:VectDropEigenTime}
& \tau_h&&\,=\,(2/\E)\,\sinh^{-1}\Bigl(\sqrt{\E h/2\gamma c}\Bigr)
&&\,\approx\, \gamma^{-1/2}\,\sqrt{2h/\E c}\,,\\
\label{eq:VectDropCoordTime}
& t_h&&\,=\,(\gamma/\E)\,\sqrt{\bigl(1+\E h/\gamma c\bigr)^2-1}
&&\,\approx\, \gamma^{+1/2}\,\sqrt{2h/\E c}\,.
\end{alignat}
\end{subequations}

This is the full solution to our problem in electrodynamics, of 
which we basically just used the Lorentz force law. It is literally 
the same in a vector theory of gravity, we just have to keep in mind 
that the `charge' $e$ is now interpreted as gravitational mass, which 
is to be set equal to the inertial mass $m$, so that $e/m=1$. Then 
$\E c$ becomes equal to the `gravitoelectric' field strength $E$, 
which directly corresponds to the strength $g$ of the scalar 
gravitational field. Having said this, we can directly compare 
(\ref{eq:VectDropTimes}) with (\ref{eq:ScalDropTimes}). For small 
field strength we see that in both cases $t_h$ is larger by a factor 
of $\gamma$ than $\tau_h$, which just reflects ordinary time dilation. 
However, unlike in the scalar case, the eigentime span $\tau_h$ also 
depends on $\gamma$ in the vector case. The independence of $\tau_h$ 
on the initial horizontal velocity is therefore a special feature of 
the scalar theory. 

\subsection{Discussion}
\label{sec:Discussion}
Let us reconsider Einstein's statements in Quote\,\ref{quote:Einstein2}, 
in which he dismisses scalar gravity for it predicting an unwanted 
dependence on the vertical acceleration on the initial horizontal 
velocity. As already noted, we do not know exactly in which formal 
context Einstein derived this result (i.e. what the ``von mir versuchten 
Theorie'' mentioned in Quote\,\ref{quote:Einstein2} actually was), 
but is seems most likely that he arrived at an equation like 
(\ref{eq:ScalThEqMotStat1}), which clearly displays the alleged 
behaviour. In any case, the diminishing effect of horizontal velocity 
onto vertical acceleration is at most of \emph{quadratic} order in $v/c$. 
\begin{Remark}
\label{rem:WhyEinsteinsConviction}
How could Einstein be so convinced that such an effect 
did not exist? Certainly there were no experiments at the time to 
support this. And yet he asserted that such a prediction ``did not 
fit with the \emph{old experience} [my italics] that all bodies 
experience the same acceleration in a gravitational field'' 
(cf. Quote\,\ref{quote:Einstein2}). What was it based on?
\end{Remark}

One way to rephrase/interpret Einstein's requirement is this: 
the time it takes for a body in free fall to drop from a height 
$h$ to the ground should be independent of its initial 
horizontal velocity. More precisely, if you drop two otherwise 
identical bodies in a static homogeneous vertical gravitational 
field at the same time from the same location, one body with 
vanishing initial velocity, the other with purely horizontal 
initial velocity, they should hit the ground simultaneously. 

But that is clearly impossible to fulfil in \emph{any} 
special-relativistic theory of gravity based on a scalar field. 
The reason is this: suppose $-\nabla^\mu\phi=(0,0,0,-g)$ is the 
gravitational field in one inertial frame. Then it takes exactly 
the same form in any other inertial frame which differs form the 
first one by 1)~spacetime translations, 2)~rotations about the 
$z$ axis, 3)~boosts in any direction within the $xy$-plane. So consider 
a situation where with respect to an inertial frame, $F$, 
body\,1 and body\,2 are simultaneously released at time $t=0$ 
from the origin, $x=y=z=0$, with initial velocities $\vec v_1=(0,0,0)$ 
and $\vec v_2=(v,0,0)$ respectively. One is interested whether the 
bodies hit the ground simultaneously. The `ground' is represented in 
spacetime by the hyperplane $z=-h$ and `hitting the ground' is taken 
to mean that the word-line of the particle in question intersects 
this hyperplane. Let another inertial frame, $F'$, move with respect 
to $F$ at speed $v$ along the $x$ axis. With respect to $F'$ both 
bodies are likewise simultaneously released at time $t'=0$ from the 
origin, $x'=y'=z'=0$, with initial velocities ${\vec v_1}'=(-v,0,0)$ 
and ${\vec v_2}'=(0,0,0)$ respectively, according to the relativistic 
law of velocity addition. The field is still static, homogeneous, 
and vertical with respect to $F'$.\footnote{\label{fnote:OliviersRemark}
This is a special feature of scalar theories. For example, in a vector 
theory, in which in $F$ the field is static homogeneous with a vertical 
electric component and no magnetic component, we would have a static 
homogeneous and vertical electric component in $F'$, but also a static 
homogeneous \emph{horizontal} magnetic component in $y$-direction.} In $F'$ 
the `ground' is defined 
by $z'=-h$, which defines the \emph{same} hyperplane in spacetime 
as $z=-h$. This is true since $F$ and $F'$ merely differ by a boost in 
$x$--direction, so that the $z$ and $z'$ coordinates coincide. Hence 
`hitting the ground' has an invariant meaning in the class of 
inertial systems considered here. However, if `hitting the ground' 
are simultaneous events in $F$ they cannot be simultaneous in $F'$ 
and vice versa, since these events differ in their $x$ coordinates. 
This leads us to the following

\begin{Remark}
\label{rem:NoSimHitGround}
Due to the usual relativity of simultaneity, the requirement of 
`hitting the ground simultaneously' cannot be fulfilled in any 
Poincar\'e invariant scalar theory of gravity.      
\end{Remark}

\noindent
But there is an obvious reinterpretation of `hitting the ground 
simultaneously', which makes perfect invariant sense in SR, 
namely the condition of `hitting the ground after the same lapse 
of eigentime'. As we have discussed in detail above, the scalar 
theory does indeed fulfil this requirement (independence of 
(\ref{eq:ScalDropEigenTime}) from $\gamma$) whereas the vector 
theory does not (dependence of (\ref{eq:VectDropEigenTime}) on 
$\gamma$). 

\begin{Remark}
\label{rem:ScalarDistinguished}
The scalar theory is distinguished by its property that the 
eigentime for free fall from a given altitude does \emph{not} 
depend on the initial horizontal velocity. 
\end{Remark}    

\noindent
In general, with regard to this requirement, the following should 
be mentioned:   

\begin{Remark}
\label{rem:Einsteins ReqNotImplied}
Einstein's requirement is (for good reasons) not implied by any of 
the modern formulations of the (weak) equivalence principle, according 
to which the worldline of a freely falling test-body (without higher 
mass-multipole-moments and without charge and spin) is determined 
by its initial spacetime point and four velocity, i.e. independent 
of the further constitution of the test body. In contrast, Einstein's 
requirement relates two motions with \emph{different} initial velocities. 
\end{Remark}

Finally we comment on Einstein's additional claim in 
Quote\,\ref{quote:Einstein2}, that there is also a similar 
dependence on the vertical acceleration on the internal energy. 
This claim, too,  does not survive closer scrutiny. Indeed, one 
might think at first that (\ref{eq:ScalThEqMotStat1}) also predicts 
that, for example, the gravitational acceleration of a box filled 
with a gas decreases as temperature increases, due to the increasing 
velocities of the gas molecules. But this arguments incorrectly neglects 
the walls of the box which gain in stress due to the rising gas 
pressure. According to (\ref{eq:FieldEq}) more stress means more 
weight. In fact, a general argument due to Laue~\cite{Laue:1911a} 
shows that these effects precisely cancel. This has been lucidly 
discussed by  Norton~\cite{Norton:1993a} and need  not be repeated here.

\section{Periapsis precession}
\label{sec:PeriapsisPrecession}
We already mentioned that the scalar theory does not predict 
any deflection of light in a gravitational field, in violation 
to experimental results. But in order to stay self contained it 
is also of interest to see directly that the system given by the 
field equation (\ref{eq:FieldEq}) and the equation of motion 
for a test particle (\ref{eq:ParticleMotion}) violates 
experimental data. This is the case if applied to planetary 
motion, more precisely to the precession of the perihelion.

Recall that the Newtonian laws of motion predict that the line 
of apsides remains fixed relative to absolute space for the motion 
of a body in a potential with $1/r$--falloff. Any deviation from 
the latter causes a rotation of the line of apsides within the 
orbital plane.  This may also be referred to as precession of 
the periapsis, the orbital point of closest approach to the centre 
of force, which is called the perihelion if the central body happens 
to be the Sun . Again we compare the result of our scalar theory with 
that of the naive scalar theory and also with that of the vector 
theory.\footnote{The scalar theory discussed by Dowker~\cite{Dowker:1965} 
was just devised to give the correct (i.e. GR-) value for the
precession of the periapsis. Since the coupling to matter is 
not discussed, it makes no statements about light deflection, 
redshift, etc.}  

There exist comprehensive treatments of periapsis precession
in various theories of gravity, like~\cite{Whitrow.Morduch:1965}. 
But rather than trying to figure out which (if any) of these 
(rather complicated) calculations apply to our theory, at least 
in a leading order approximation, it turns out to be easier, more 
instructive, and mathematically more transparent to do these 
calculations from scratch. A convenient way to compute the 
periapsis precession in perturbed $1/r$--potentials is provided 
by the following proposition, which establishes a convenient 
and powerful technique for calculating the periapsis precession 
in a large variety of cases. 
\begin{proposition}
\label{prop:LL-PeriAdvFormula}
Consider the Newtonian equations of motion for a test particle 
of mass $m$ in a perturbed Newtonian potential
\begin{equation}
\label{eq:PertNewtonPot}
U(r)=-\,\frac{\alpha}{r}+\Delta U(r)\,,
\end{equation}
where $\alpha>0$ and $\Delta U(r)$ is the perturbation.
The potential is normalised so that it tends to zero 
at infinity, i.e. $\Delta U(r\rightarrow\infty)\rightarrow 0$. 
Let $2\pi+\Delta\varphi$ 
denote the increase of the polar angle between two successive 
occurrences of periapsis. Hence $\Delta\varphi$ represents the 
excess over a full turn, also called the `periapsis shift per 
revolution'. Then the first-order contribution of $\Delta U$ to 
$\Delta\varphi$ is given by 
\begin{equation}
\label{eq:LL-PeriAdvFormula}
\Delta\varphi=
\frac{\partial}{\partial L}
\left\{\frac{2m}{L}\int_0^\pi r_*^2(\varphi;L,E)\
\Delta U\bigl(r_*(\varphi;L,E)\bigr)\ d\varphi\right\}\,.
\end{equation}
Here $\varphi\mapsto r_*(\varphi;L,E)$ is the solution of the 
unperturbed problem (Kepler orbit) with angular momentum $L$
and energy $E$. (As we are interested in bound orbits, we 
have $E<0$.) It is given by 
\begin{subequations}
\label{eq:KeplerOrbit}
\begin{equation}
\label{eq:KeplerOrbit1}
r_*(\varphi;L,E)=\frac{p}{1+\varepsilon\,\cos\varphi}\,,
\end{equation}
where
\begin{alignat}{2}
\label{eq:KeplerOrbit2}
& p:&&\,=\,\frac{L^2}{m\alpha}\,,\\
\label{eq:KeplerOrbit3}
& \varepsilon:&&\,=\,\sqrt{1+\frac{2EL^2}{m\alpha^2}}\,.
\end{alignat}
\end{subequations}
Note that the expression in curly brackets on the right hand side 
of (\ref{eq:LL-PeriAdvFormula}) is understood as function 
of $L$ and $E$, so that the partial differentiation is to be 
taken at constant $E$. 
\end{proposition}
\begin{proof}
In the Newtonian setting, the conserved quantities of 
energy and angular momentum for the motion in a plane 
coordinatised by polar coordinates, are given by  
\label{eq:NewtonianEnergyAngMom}
\begin{alignat}{2}
\label{eq:NewtonianEnergy1}
&E&&\,=\,\tfrac{1}{2}m(r'^2+r^2\varphi'^2)+U(r)\,,\\
\label{eq:NewtonianAngMom}
&L&&\,=\,mr^2\varphi'\,,
\end{alignat}
where a prime represents a $t$-derivative.  
Eliminating $\varphi'$ in (\ref{eq:NewtonianEnergy1})
via (\ref{eq:NewtonianAngMom}) and also using 
(\ref{eq:NewtonianAngMom}) to re-express $t$-derivatives in 
terms of $\varphi$-derivatives, we get
\begin{equation}
\label{eq:NewtonianEnergy2}
\frac{L^2}{m^2r^4}\left((dr/d\varphi)^2+r^2\right)
=2\,\frac{E-U}{m}\,.
\end{equation}
This can also be written in differential form,
\begin{equation}
\label{eq:NewtonianEnergyAlt}
d\varphi=\frac{\pm dr\,L/r^2}{\sqrt{2m\bigl(E-U(r)\bigr)-L^2/r^2}}\,,
\end{equation}  
whose integral is just given by (\ref{eq:KeplerOrbit}). 

Now, the angular change between two successive occurrences of periapsis 
is twice the angular change between periapsis, $r_{\sss\rm min}$,
and apoapsis, $r_{\sss\rm max}$:
\begin{equation}
\label{eq:NewtonianAngularShift}
\begin{split}
\Delta\varphi+2\pi&=2\int_{r_{\sss\rm min}}^{r_{\sss\rm max}}
\frac{dr\,L/r^2}{\sqrt{2m\bigl(E-U(r)\bigr)-L^2/r^2}}\\
&=-2\,\frac{\partial}{\partial L}\left\{
\int_{r_{\sss\rm min}}^{r_{\sss\rm max}}dr\;
\sqrt{2m\bigl(E-U(r)\bigr)-L^2/r^2}\right\}\,,\\
\end{split}
\end{equation}
where the term in curly brackets is considered as function of $L$ and 
$E$ and the partial derivative is for constant $E$. 

Formula (\ref{eq:NewtonianAngularShift}) is exact. Its sought-after 
approximation is obtained by writing $U(r)=-\alpha/r+\Delta U(r)$ and 
expanding the integrand to linear order in $\Delta U$. Taking into 
account that the zeroth order term just cancels the $2\pi$ on the left 
hand side, we get: 
\begin{equation}
\label{eq:NewtonianAngularShiftApprox}
\begin{split}
\Delta\varphi
&\approx \frac{\partial}{\partial L}
\left\{2m\int_{r_{\sss\rm min}}^{r_{\sss\rm max}}
\frac{\Delta U(r)\,dr}{\sqrt{2m\bigl(E+\alpha/r\bigr)-L^2/r^2}}
\right\}\\
&\approx \frac{\partial}{\partial L}
\left\{\frac{2m}{L}\int_0^\pi r_*^2(\varphi;L,E)\,
\Delta U\bigl(r_*(\varphi;L,E)\bigr)\,d\varphi\right\}\,.\\
\end{split}
\end{equation}
In the second step we converted the $r$--integration into an integration 
over the azimuthal angle $\varphi$. This we achieved by making use of 
the identity that one obtains from (\ref{eq:NewtonianEnergyAlt})
with $U(r)=-\alpha/r$ and $r$ set equal to the Keplerian solution 
curve $r_*(\varphi;L,M)$ for the given parameters $L$ 
and $E$. Accordingly, we replaced the integral limits $r_{\sss\rm min}$ 
and $r_{\sss\rm max}$ by the corresponding angles $\varphi=0$ 
and $\varphi=\pi+\Delta\varphi/2$ respectively. Since the integrand is 
already of order $\Delta U$, we were allowed to replace the upper limit 
by $\varphi=\pi$, so that the integral limits now correspond to the 
angles for the minimal and maximal radius of the 
unperturbed Kepler orbit $r_*(\varphi;L,E)$ given by 
(\ref{eq:KeplerOrbit1}).      
\end{proof}

Let us apply this proposition to the general class of cases where 
$\Delta U=\Delta_2U+\Delta_3 U$ with 
\begin{subequations}
\label{eq:PotentialPerturb}
\begin{alignat}{2}
\label{eq:PotentialPerturb1}
&\Delta_2U(r)&&\,=\,\delta_2/r^2\,,\\
\label{eq:PotentialPerturb2}
&\Delta_3U(r)&&\,=\,\delta_3/r^3\,.
\end{alignat}
\end{subequations} 
In the present linear approximation in $\Delta U$
the effects of both perturbations to $\Delta\varphi$ 
simply add, so that $\Delta\varphi=\Delta_2\varphi+\Delta_3\varphi$. 
The contributions $\Delta_2\varphi$ and $\Delta_3\varphi$
are very easy to calculate from 
(\ref{eq:LL-PeriAdvFormula}). The integrals are 
trivial and give $\pi\delta_2$ and $\pi\delta_3/p$ 
respectively. Using (\ref{eq:KeplerOrbit2}) in the second case 
to express $p$ as function of $L$, then doing the $L$-differentiation,
and finally eliminating $L$ again in favour of $p$ using  
(\ref{eq:KeplerOrbit2}), we get 
\begin{subequations}
\label{eq:PeriastronAdvanceFormula}
\begin{alignat}{3}
\label{eq:PeriastronAdvanceFormula1}
&\Delta_2\varphi
&&\,=\,-\,2\pi\,\left[\frac{\delta_2/\alpha}{p}\right]
&&\,=\,-\,2\pi\,\left[\frac{\delta_2/\alpha}{a(1-\varepsilon^2)}\right]\,,\\
\label{eq:PeriastronAdvanceFormula2}
&\Delta_3\varphi
&&\,=\,-\,6\pi\,\left[\frac{\delta_3/\alpha}{p^2}\right]
&&\,=\,-\,6\pi\,\left[\frac{\delta_3/\alpha}{a^2(1-\varepsilon^2)^2}\right]\,,
\end{alignat}
\end{subequations}
were we also expressed $p$ in terms of the semi-major axis 
$a$ and the eccentricity $\varepsilon$ via $p=a(1-\varepsilon^2)$,
as it is usually done. Clearly this method allows to calculate 
in a straightforward manner the periapsis shifts for 
general perturbations $\Delta_nU=\delta_n/r^n$. For example,
the case $n=3$ is related to the contribution from the
quadrupole moment of the central body.     

\subsection{Scalar model-theory}
\label{sec:PeriPressScalarModel}
All this applies directly to the scalar theory if its equation 
of motion is written in the Newtonian form (\ref{eq:ScalThEqMotStat2}).
The static and rotationally symmetric solution to 
(\ref{eq:FieldEq}) outside the point source is $\Phi(r)=-GM/r$,
so that
\begin{equation}
\label{eq:ScalThPointMassSol}
\tilde\phi(r)=
(mc^2/2)\gamma_0^{-2}\left(1-\frac{GM}{rc^2}\right)^2\,.
\end{equation}
In order to normalize the potential so that it assumes the value 
zero at spatial infinity we just need to drop the constant term. 
This leads to to  
\begin{subequations}
\label{eq:ScalThPointMassExpCoef}
\begin{alignat}{2}
\label{eq:ScalThPointMassExpCoef1}
&\alpha&&\,=\,\gamma_0^{-2}GMm\,,\\
\label{eq:ScalThPointMassExpCoef2}
&\delta_2&&\,=\alpha\,\frac{GM}{2c^2}\,,
\end{alignat}
\end{subequations}
so that 
\begin{equation}
\label{eq:ScalThPeriastronShift}
\Delta\varphi=\Delta_2\varphi=
-\,\pi\,\left[\frac{GM/c^2}{a(1-\varepsilon^2)}\right]
=-\,\tfrac{1}{6}\Delta_{\sss\rm GR}\varphi\,,
\end{equation}
where $\Delta_{\sss\rm GR}\varphi$ is the value predicted by 
GR. Hence scalar gravity leads to a \emph{retrograde}
periapsis precession

\subsection{Naive scalar theory}
\label{sec:PeriPressNaiveScalar}
In the naive scalar theory we have $\phi(r)=-GM/r$ in 
(\ref{eq:ScalThEqMotStat2}) and therefore  
\begin{equation}
\label{eq:NaiveScalThPointMassSol}
\begin{split}
\tilde\phi(r)&=
(mc^2/2)\gamma_0^{-2}\exp\bigl(-2GM/c^2r\bigr)\\
&=(mc^2/2)\gamma_0^{-2}\left\{1
-2\left(\frac{GM}{c^2r}\right)
+2\left(\frac{GM}{c^2r}\right)^2
-\frac{4}{3}\,\left(\frac{GM}{c^2r}\right)^3+\cdots\right\}
\,.\\
\end{split}
\end{equation}
Again we subtract the constant term to normalize the potential 
so as to assume the value zero at infinity. Then we simply read 
off the coefficients $\alpha,\delta_2$, and $\delta_3$: 
\begin{subequations}
\label{eq:NaiveScalThPointMassExpCoef}
\begin{alignat}{5}
\label{eq:NaiveScalThPointMassExpCoef1}
&\alpha  &&\,=\,&&2&&(GM/c^2)  \ &&(mc^2/2)\gamma_0^{-2}\,,\\
\label{eq:NaiveScalThPointMassExpCoef2}
&\delta_2&&\,=\,&&2&&(GM/c^2)^2\ &&(mc^2/2)\gamma_0^{-2}\,,\\
 \label{eq:NaiveScalThPointMassExpCoef3}
&\delta_3&&\,=\,-&&\tfrac{4}{3}&&(GM/c^2)^3\ &&(mc^2/2)\gamma_0^{-2}\,.
\end{alignat}
\end{subequations}
Hence we have
\begin{subequations}
\label{eq:NaiveScalThPeriastronShift}
\begin{equation}
\label{eq:NaiveScalThPeriastronShift1}
\Delta\varphi=\Delta_2\varphi+\Delta_3\varphi\,,
\end{equation}
where
\begin{alignat}{3}
\label{eq:NaiveScalThPeriastronShift2}
&\Delta_2\varphi&&\,=\,-\,2\pi\,
&&\left[\frac{GM/c^2}{a(1-\varepsilon^2)}\right]\,,\\
\label{eq:NaiveScalThPeriastronShift3}
&\Delta_3\varphi&&\,=\,+\,4\pi\,
&&\left[\frac{GM/c^2}{a(1-\varepsilon^2)}\right]^2\,.
\end{alignat}
\end{subequations}

Recall that (\ref{eq:PeriastronAdvanceFormula}) neglects quadratic 
and higher order terms in $\Delta U$. If we expand $\Delta U$ in 
powers of $GM/c^2r$, as done in (\ref{eq:NaiveScalThPointMassSol}),
it would be inconsistent to go further than to third order because 
$\Delta U$ starts with the quadratic term so that the neglected 
corrections of order $(\Delta U)^2$ start with fourth powers in 
$GM/c^2r$. Hence (\ref{eq:NaiveScalThPeriastronShift}) gives the 
optimal accuracy obtainable with (\ref{eq:LL-PeriAdvFormula}). 
For solar-system applications $GM/c^2a$ is of the order of $10^{-8}$
so that the quadratic term (\ref{eq:NaiveScalThPeriastronShift3}) can 
be safely neglected. Comparison of (\ref{eq:NaiveScalThPeriastronShift2})
with (\ref{eq:ScalThPeriastronShift}) shows that the naive scalar 
theory gives a value twice as large as that of the consistent 
model-theory, that is, $-1/3$ times the correct value (predicted by GR). 

\subsection{Vector theory }
We start from the following 
\begin{proposition}
The equations of motion (\ref{eq:LorentzEqMotion})
for a purely `electric' field, where $F_{0i}=-F_{i0}=E_i/c$ 
and all other components of $F_{\mu\nu}$ vanish, is equivalent 
to
\begin{equation}
\label{eq:EqMotVectElec}
\bigl(\gamma(t)\vec x'(t)\bigr)'=c\vec\E\bigl(\vec x(t)\bigr)\,,
\end{equation}
where again the prime $'$ denotes $d/dt$, 
$\gamma(t):=1/\sqrt{1-\Vert\vec x'(t)\Vert^2/c^2}$, and
$\vec\E:=e\vec E/mc$.
\end{proposition}
\begin{proof}
We have $d/ds=\gamma\,d/dt$, 
$d\gamma/dt=\gamma^3(\vec\beta\cdot\vec\beta')$. Now,
\begin{equation}
\label{eq:EqMotVectElecProp1}
\ddot z^\mu=
c\gamma\,\bigl(\gamma',(\gamma\vec\beta)'\bigr)\,,
\quad\mathrm{and}\quad
(e/m)\,F^\mu_\nu\dot z^\nu=
c\gamma\,\bigl(\vec\E\cdot\beta,\vec\E\bigr)\,,
\end{equation}
so that (\ref{eq:LorentzEqMotion}) is equivalent to 
\begin{subequations}
\label{eq:LorentzEqMotionSplit}
\begin{alignat}{2}
\label{eq:LorentzEqMotionSplit1}
\vec\E\cdot\vec\beta&\,=\,\gamma'&\,=\,&
\gamma^3(\vec\beta\cdot\vec\beta')\,,\\
\label{eq:LorentzEqMotionSplit2}
\vec\E&\,=\,(\gamma\vec\beta)'&\,=\,&
\gamma^3\vec\beta'_{\Vert}+\gamma\vec\beta'_\perp\,,
\end{alignat}
\end{subequations}
where $\Vert$ and $\perp$ refer to the projections parallel 
and perpendicular to $\vec\beta$ respectively. Since 
(\ref{eq:LorentzEqMotionSplit2}) implies  
(\ref{eq:LorentzEqMotionSplit1}), (\ref{eq:LorentzEqMotion})
is equivalent to the former. 
\end{proof}

We apply this to a spherically symmetric field, where 
$c\vec\E=-\vec\nabla\phi$ with $\phi(r)=-GM/r$. 
This implies conservation of angular momentum, the modulus of 
which is now given by 
\begin{equation}
\label{eq:VectConsAngMom}
L=\gamma\,m r^2\varphi'\,.
\end{equation}
Note the explicit appearance of $\gamma$, which, e.g., is not present 
in the scalar case, as one immediately infers from 
(\ref{eq:ScalThEqMotStat1}). This fact makes 
Proposition\,\ref{prop:LL-PeriAdvFormula} not immediately applicable. 
We proceed as follows: scalar multiplication of 
(\ref{eq:EqMotVectElec}) with $\vec v=\vec x'$ and $m$ leads to the 
following expression for the conserved energy: 
\begin{equation}
\label{eq:VectConsEnergy} 
E=mc^2(\gamma-1)+U\,,
\end{equation}
where $U=m\phi$. This we write in the form
\begin{subequations}
\begin{equation}
\label{eq:VectConsGammaSquared1} 
\gamma^2=\left(1+\frac{E-U}{mc^2}\right)^2\,.
\end{equation}
On the other hand, we have
\begin{equation}
\label{eq:VectConsGammaSquared2} 
\gamma^2\equiv 1+(\beta\gamma)^2=1+(\gamma/c)^2(r'^2+r^2\varphi'^2)=
1+\frac{L^2}{m^2c^2r^4}\left((dr/d\varphi)^2+r^2\right)\,,
\end{equation}
\end{subequations}
where we used (\ref{eq:VectConsAngMom}) to eliminate $\varphi'$ 
and convert $r'$ into $dr/d\varphi$, which also led to a cancellation 
of the factors of $\gamma$. Equating (\ref{eq:VectConsGammaSquared1}) 
and (\ref{eq:VectConsGammaSquared2}), we get
\begin{equation}
\label{eq:VectThNewtonianForm1} 
\frac{L^2}{m^2r^4}\left((dr/d\varphi)^2+r^2\right)=
2\frac{\tilde E-\tilde U}{m}
\end{equation}
where 
\begin{subequations}
\label{eq:VectThNewtonianForm2}
\begin{alignat}{2}
\label{eq:VectThNewtonianForm2a}
& \tilde E&&\,:=\,E\bigl(1+E/2mc^2\bigr)\,,\\
\label{eq:VectThNewtonianForm2b}
& \tilde U&&\,:=\,U\bigl(1+E/mc^2\bigr)-U^2/2mc^2\,.
\end{alignat}
\end{subequations}
Equation (\ref{eq:VectThNewtonianForm1}) is just of the form 
(\ref{eq:NewtonianEnergy2}) with $\tilde E$ and $\tilde U$ 
replacing $E$ and $U$. In particular we have for 
$U=m\phi=-GMm/r$:
\begin{equation}
\label{eq:VectThPotentialCorr1}
\tilde U(r)=-\frac{\alpha}{r}+\frac{\delta_2}{r^2}\,,
\end{equation}
with 
\begin{subequations}
\label{eq:VectThPotentialCorr2}
\begin{alignat}{2}
&\alpha&&\,=\,GMm(1+E/mc^2)\,,\\
&\delta_2&&\,=\,-\,\frac{G^2M^2m}{2c^2}\,.
\end{alignat}
\end{subequations}
In leading approximation for small $E/mc^2$ we have 
$\delta_2/\alpha\approx-GM/2c^2$. The advance of the periapsis  
per revolution can now be simply read off 
(\ref{eq:PeriastronAdvanceFormula1}):
\begin{equation}
\label{eq:VectThPeriAdv}
\Delta\varphi=\pi\,\left[\frac{GM/c^2}{a(1-\varepsilon^2)}\right]
=\tfrac{1}{6}\Delta_{\sss\rm GR}\varphi\,.
\end{equation}
This is the same amount as in the scalar model-theory 
(compare (\ref{eq:ScalThPeriastronShift})) but of opposite 
sign, corresponding to a \emph{prograde} periapsis precession 
of 1/6 the value predicted by GR.   
 
\section{Energy conservation}
In this section we finally turn to Einsteins argument of the 
Entwurf Paper concerning energy conservation. From a modern 
viewpoint, Einstein's claim of the violation of energy 
conservation seems to fly in the face of the very concept of 
Poincar\'e invariance. After all, time translations are among 
the symmetries of the Poincar\'e group, thus giving rise to a 
corresponding conserved Noether charge. Its conservation is a 
theorem and cannot be questioned. The only thing that seems 
logically questionable is whether this quantity does indeed 
represent physical energy. So how could Einstein arrive at 
his conclusion?    

\subsection{Einstein's argument}
\begin{center}
\begin{figure}
\hspace{1.0cm}
\begin{minipage}[b]{0.25\linewidth}
\centering\epsfig{figure=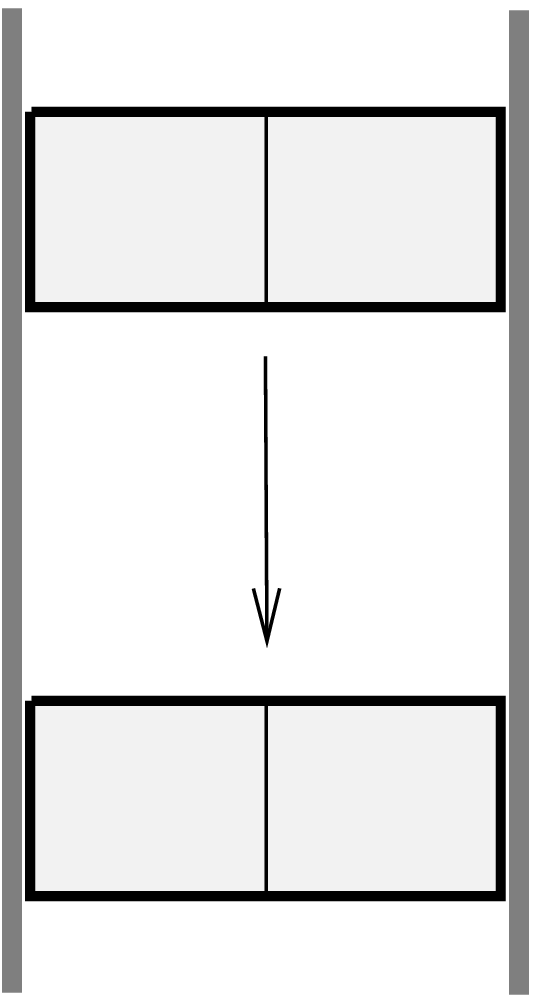,width=\linewidth}
\caption{\label{fig:LoweringBox}
Lowering the box in the gravitational field with side walls attached.}
\end{minipage}\hfill
\begin{minipage}[b]{0.28\linewidth}
\centering\epsfig{figure=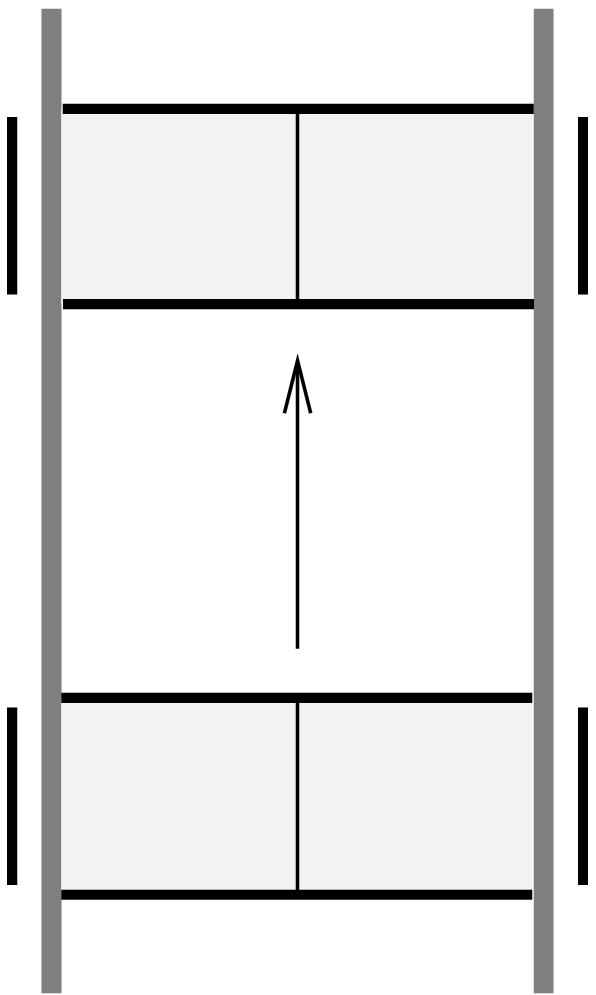,width=\linewidth}
\caption{\label{fig:RaisingBox} 
Raising the box in the gravitational field with side walls 
taken off.}
\end{minipage}
\hspace{1.0cm}
\end{figure}
\end{center}

Einstein first pointed out that the source for the gravitational
field must be a scalar built from the matter quantities alone, and 
that the only such scalar is the trace $T^{\mu}_{\mu}$ of the 
energy-momentum tensor (as pointed out to Einstein by Laue, 
as Einstein acknowledges, calling  $T^{\mu}_{\mu}$ the ``Laue Scalar''). 
Moreover, for \emph{closed stationary systems}, the so-called 
Laue-Theorem~\cite{Laue:1911a} for static systems (later slightly 
generalised to stationary ones) states that the space integral of 
$T^{\mu\nu}$ must vanish, except for $\mu=0=\nu$; hence the space 
integral of $T^{\mu}_{\mu}$ equals 
that of $T^{00}$, which means that the total (active and passive) 
gravitational mass of a closed stationary system equals its inertial 
mass. However, if the system is not closed, the weight depends on 
the stresses (the spatial components $T^{ij}$).
\begin{wrapfigure}{r}{0.3\linewidth}
\vspace{-0.3cm}
\centering\epsfig{figure=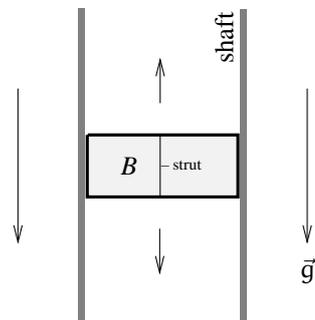,width=\linewidth}
\put(-5,17){\small $\vec g$}
\put(-57,58){\tiny -- strut}
\put(-72,56){\small\textit{B}}
\put(-30,98){\begin{rotate}{90}\small shaft\end{rotate}}
\caption{\label{fig:SlidingBox}
Sliding box filled with radiation in a gravitational field $\vec g$.}
\end{wrapfigure}
His argument proper is then as follows (compare 
Fig.\,\ref{fig:SlidingBox}): consider a box, $B$, filled 
with electromagnetic radiation of total energy $E$. We idealise the walls 
of the box to be inwardly perfectly mirrored and of infinite stiffness,
so as to be able to support normal stresses (pressure) without suffering 
any deformation. The box has an additional vertical strut in the middle, 
connecting top and bottom walls, which supports all the vertical material 
stresses that counterbalance the radiation pressure, so that the 
side walls merely sustain normal and no tangential stresses. 
The box can slide without friction along a vertical shaft whose 
cross section corresponds exactly to that of the box. The walls of 
the shaft are likewise idealised to be inwardly perfectly mirrored 
and of infinite stiffness. The whole system of shaft and box is 
finally placed in a homogeneous static gravitational field, $\vec g$,
which points vertically downward. Now we perform the following 
process. We start with the box being placed in the shaft in the upper 
position. Then we slide it down to the lower position; see 
Fig.\,\ref{fig:LoweringBox}. There we remove the side walls of the 
box---without any radiation leaking out---such that the sideways
pointing pressures are now provided by the shaft walls. The strut 
in the middle is left in position to further support all the vertical 
stresses, as 
before. Then the box together with the detached side walls are pulled 
up to their original positions; see Fig.\,\ref{fig:RaisingBox}. 
Finally the system is reassembled so that it assumes its initial state. 
Einstein's claim is now that in a very general class of imaginable 
scalar theories the process of pulling up the parts needs less work 
than what is gained in energy in letting the box (with side walls 
attached) down. Hence he concluded that such theories necessarily 
violate energy conservation.

Indeed, radiation-plus-box is a closed stationary system in 
Laue's sense. Hence the weight 
of the total system is proportional to its total energy, $E$, which we 
may pretend to be given by the radiation energy alone since the 
contributions from the rest masses of the walls will cancel in the 
final energy balance, so that we may formally set them to zero at 
this point. Lowering this box by an amount $h$ in a static homogeneous 
gravitational field of strength $g$ results in an energy gain of 
$\Delta E=hgE/c^2$. So despite the fact that radiation has a traceless 
energy-momentum tensor, \emph{trapped} radiation has a weight given by 
$E/c^2$. This is due to the radiation pressure which puts the walls 
of the trapping box under tension. For each parallel pair of side-walls 
the tension is just the radiation pressure, which is one-third of the 
energy density. So each pair of side-walls contribute $E/3c^2$ to 
the (passive) gravitational mass (over and above their rest mass, 
which we set to zero) in the lowering process when stressed, and zero in the 
raising process when unstressed. Hence, Einstein concluded, there 
is a net gain in energy of $2E/3c^3$ (there are two pairs of side 
walls). 

But it seems that Einstein neglects a crucial contribution to 
the energy balance. In contrast to the lowering process, the 
state of the shaft \emph{is} changed during the lifting process,
and it is this additional contribution which just renders Einstein's 
argument inconclusive. Indeed, when the side walls are first removed 
in the lower position, the walls of the shaft necessarily come under 
stress because they now need to provide the horizontal balancing 
pressures. In the raising process that stress distribution of the 
shaft is translated upwards. But that \emph{does} cost energy in the 
theory discussed here, even though it is not associated with any 
proper transport of the material the shaft is made from. As already 
pointed out, stresses make their own contribution to weight, 
independent of the nature of the material that supports them. 
In particular, a redistribution of stresses in a material immersed 
in a gravitational field generally makes a non-vanishing 
contribution to the energy balance, even if the material does not 
move. This will be seen explicitly below. There seems to be 
only one paper which explicitly expresses some uneasiness with 
Einstein's argument, due to the negligence of ``edge effects'' 
(\cite{Wellner.Sandri:1964}, p.\,37), however without going into 
any details, letting alone establishing energy expressions and 
corresponding balance equations.

\subsection{Energy conservation in the scalar model-theory}
There are 10 conserved currents corresponding to Poincar\'e-invariance. 
In particular, the total energy $E$ relative to an inertial system
is conserved. For a particle coupled to gravity it is easily 
calculated and consists of three contributions corresponding to 
the gravitational field, the particle, and the interaction-energy  
shared by the particle and the field:
\begin{subequations}
\label{eq:Energies}
\begin{alignat}{2}
\label{eq:EnergiesGrav}
& E_{\sss\rm gravity}
&&\,=\,\frac{1}{2\kappa c^2}\int d^3x\ 
  \bigl((\partial_{ct}\Phi)^2+(\vec\nabla\Phi)^2\bigr)\,,\\
\label{eq:EnergiesPart}
& E_{\sss\rm particle} 
&&\,=\, mc^2\,\gamma(v)\,,\\
\label{eq:EnergiesInt}
& E_{\sss\rm interaction} 
&&\,=\, m\,\gamma(v)\,\Phi\bigl(\vec z(t),t\bigr)\,.
\end{alignat}
\end{subequations}

Let us return to general matter models and let $T_{\sss\rm total}^{\mu\nu}$ 
be the total stress-energy tensor of the gravity-matter-system. 
It is the sum of three contributions:
\begin{equation}
T_{\sss\rm total}^{\mu\nu}=
T_{\sss\rm gravity}^{\mu\nu}+
T_{\sss\rm matter}^{\mu\nu}+
T_{\sss\rm interaction}^{\mu\nu}\,,
\label{total-setensor}
\end{equation}
where\footnote{We simply use the standard expression for the 
canonical energy-momentum tensor, which is good enough in the 
present case. If $S=\int L\,dtd^3x$, it is given by 
$T^\mu_\nu:=(\partial L/\partial\Phi_{,\mu})\Phi_{,\nu}-\delta^\mu_\nu L$,
which here (generally for scalar fields) gives rise to a symmetric tensor, 
$T^{\mu\nu}=T^{\nu\mu}$.}
\begin{subequations}
\label{eq:EM-Tensors}
\begin{alignat}{2}
\label{eq:EM-TensorsGrav}
& T_{\sss\rm gravity}^{\mu\nu}
&&\,=\, \frac{1}{\kappa c^2}\bigl(\partial^{\mu}\Phi\partial^{\nu}\Phi
     -\tfrac{1}{2}\eta^{\mu\nu}\partial_{\lambda}\Phi
      \partial^{\lambda}\Phi\bigr)\,,\\
\label{eq:EM-TensorsMatter}
& T_{\sss\rm matter}^{\mu\nu}
&&\,=\,\text{depending on matter model}\,,\\
\label{eq:EM-TensorsInt}
&T_{\sss\rm interaction}^{\mu\nu}
&&\,=\,\eta^{\mu\nu}(\Phi/c^2) T_{\sss\rm matter}\,.
\end{alignat}
\end{subequations}
Energy-momentum-conservation is expressed by
\begin{equation}
\partial_{\mu}{T_{\sss\rm total}}^{\mu\nu}=F^{\nu}_{\sss\rm external}\,,
\label{em-conservation1}
\end{equation}
where $F^{\nu}_{\sss\rm external}$ is the four-force of a possible
\emph{external} agent. The 0-component of it (i.e. energy conservation)
can be rewritten in the form 
\begin{equation}
\mbox{external power supplied}\ =\
\frac{d}{dt}\int_D d^3x\ {T^{00}_{\sss\rm total}}+
\int_{\partial D}{T^{0k}_{\sss\rm total}}n_k\ d\Omega\,,
\label{em-conservation2} 
\end{equation}
for any bounded spatial region $D$. If the matter system is itself 
of finite spatial extent, meaning that outside some bounded spatial 
region, $D$, $T_{\sss\rm matter}^{\mu\nu}$ vanishes identically, and 
if we further assume that no gravitational radiation escapes to infinity, 
the  surface integral in (\ref{em-conservation2}) vanishes identically. 
Integrating (\ref{em-conservation2}) over time we then get 
\begin{equation}
\mbox{external energy supplied}\ =\ \Delta E_{\sss\rm gravity}
+\Delta E_{\sss\rm matter} +\Delta E_{\sss\rm interaction}\,,
\label{em-conservation3}
\end{equation}
with
\begin{equation} 
E_{\sss\rm interaction}=
\int_Dd^3x\ (\Phi/c^2) T_{\sss\rm matter}\,,
\label{interaction-energy}
\end{equation}
and where $\Delta(\mbox{something})$ denotes the difference 
between the initial and final value of $\mbox{`something'}$.
If we apply this to a process that leaves the \emph{internal} 
energies of the gravitational field and the matter system unchanged,
for example a processes where the matter system, or at least the relevant 
parts of it, are \emph{rigidly} moved in the gravitational field, like 
in Einstein's Gedankenexperiment of the `radiation-shaft-system', 
we get
\begin{equation}
\mbox{external energy supplied}\ =\ \Delta\left\{
\int_Dd^3x\ (\Phi/c^2) T_{\sss\rm matter}\right\}\,.
\label{em-conservation4}
\end{equation}
Now, my understanding of what a valid claim of energy 
non-conservation in the present context would be is to 
show that \emph{this} equation can be violated. But this 
is \emph{not} what Einstein did (compare Conclusions).

If the matter system stretches out to infinity and conducts 
energy and momentum to infinity, then the surface term that was 
neglected above gives a non-zero contribution that must be included 
in (\ref{em-conservation4}). Then a proof of violation of energy 
conservation must disprove this modified equation. 
(Energy conduction to infinity as such is not in any disagreement
with energy conservation; you have to prove that they do not balance
in the form predicted by the theory.)

\subsection{Discussion}
For the discussion of Einstein's Gedankenexperiment the term 
(\ref{interaction-energy}) is the relevant one. It accounts for 
the \emph{weight of stress}. Pulling up a radiation-filled box inside 
a shaft also moves up the stresses in the shaft walls that must 
act sideways to balance the radiation pressure. This lifting of 
stresses to higher gravitational potential costs energy, according to 
the theory presented here. This energy was neglected by Einstein,
apparently because it is not associated with a transport of matter. 
He included it in the lowering phase, where the side-walls of the box
are attached to the box and move with it, but neglected them in the 
raising phase, where the side walls are replaced by the shaft, which 
does not move. But as far as the `weight of stresses' is concerned, 
this difference is irrelevant. What (\ref{interaction-energy}) tells 
us is that raising stresses in an ambient gravitational potential 
costs energy, irrespectively of whether it is associated with an 
actual transport of the stressed matter or not. This would be just 
the same for the transport of heat in a heat-conducting material. 
Raising the heat distribution against the gravitational field costs 
energy, even if the material itself does not move.

\section{Conclusion}
\label{sec:Conclusion}
From the foregoing I conclude that, taken on face value, neither 
of Einstein's reasonings that led him to dismiss scalar theories 
of gravity prior to being checked against experiments are convincing. 
First, energy---as defined by Noether's theorem---\emph{is} conserved 
in our model-theory. Note also that the energy of the free 
gravitational field is positive definite in this theory. 
Second, the \emph{eigen}time for 
free fall in a homogeneous static gravitational field \emph{is} 
independent of the initial horizontal velocity. Hence our model-theory 
serves as an example of an internally consistent theory which, 
however, is experimentally ruled out. As we have seen, it predicts 
$-1/6$ times the right perihelion advance of Mercury and also no 
light deflection (not to mention Shapiro time-delay, gravitational 
red-shift, as well as other accurately measured effects which are 
correctly described by GR). 

The situation is slightly different in a special-relativistic 
vector theory of gravity (Spin\,1, mass\,0). Here the energy is clearly 
still conserved (as in any Poincar\'e invariant theory), but the 
energy of the radiation field is negative definite due to a sign change 
in Maxwell's equations which is necessary to make like charges (i.e. 
masses) attract rather than repel each other. Hence there 
exist runaway solutions in which a massive particle self-accelerates 
unboundedly by radiating negative gravitational radiation. Also, 
the free-fall eigentime now does depend on the horizontal velocity,
as we have seen. Hence, concerning these theoretical aspects, scalar 
gravity is much better behaved.  

This leaves the question unanswered why Einstein thought it 
necessary to give up the identification of Minkowski geometry 
with the physical geometry, as directly measured with physical 
clocks and rods (cf. the discussion at the end of 
Section\,\ref{sec:HistBackground}). Einstein made it sound as if 
this was the only way to save energy conservation. This, as we 
have seen, is not true. But there may well be other reasons to 
contemplate more general geometries than that of Minkowski space
from considerations of scalar gravity as presented here, merely 
by looking at the gravitational interaction of models 
for `clocks' and `rods'. A simple such model would be given by 
an electromagnetically bound system, like an atom, where (classically 
speaking) an electron orbits a charged nucleus (both modelled as 
point masses). Place this system in a 
gravitational field that varies negligibly over the spatial extent 
of the atom and over the time of observation. The electromagnetic 
field produced by the charges will be unaffected by the gravitational 
field (due to its traceless energy momentum tensor). However, 
(\ref{eq:ScalGrav4}) tells us that the dynamics of the particle 
is influenced by the gravitational field. The effect can be 
conveniently summarised by saying that the masses of point 
particles scale by a factor of $1+\Phi/c^2=\exp(\phi/c^2)$
when placed in the potential $\phi$.  This carries over to 
Quantum Mechanics so that atomic length scales, like the Bohr radius
(in MKSA units)
\begin{equation}
\label{eq:DefBohrRadius}
a_0:=\frac{\varepsilon_0\,h^2}{m\,\pi e^2}\,,
\end{equation}
and time scales, like the Rydberg period (inverse Rydberg frequency)
\begin{equation}
\label{eq:DefRydbergPeriod}
T_R:=\frac{8\varepsilon_0^2h^3}{me^4}\,,
\end{equation}
change by a factor $\exp(-\phi/c^2)$ due to their inverse proportionality
to the electron mass $m$ ($h$ is Planck's constant, $e$ 
the electron charge, and $\varepsilon_0$ the vacuum permittivity). 
This means that, relative to the units on which the Minkowski metric is 
based, atomic units of length and time vary in a way depending on 
the potential. Transporting the atom to a spacetime position in which 
the gravitational potential differs by an amount $\Delta\phi$ 
results in a diminishment (if $\Delta\phi>0$) or enlargement 
(if $\Delta\phi<0$) of its size and period 
\emph{relative to Minkowskian units}. This effect is universal for all 
atoms. 

The question then arises as to the physical significance of 
the Minkowski metric. Should we not rather \emph{define} spacetime 
lengths by what is measured using atoms? After all, as Einstein 
repeatedly remarked, physical notions of spatial lengths and times 
should be based on physically constructed rods and clocks which are 
consistent with our dynamical equations. The Minkowski metric 
would then merely turn into a redundant structure with no 
\emph{direct} observational significance.\footnote{
Note that we are extrapolating here, since our argument is based 
on the Lagrangian for the specific context. We have not shown 
that the Minkowski metric can be eliminated in general. `Clocks' 
and `rods' not based on atomic frequencies and lengths scales are 
clearly conceivable. Moreover, epistemologically speaking, 
eliminability in a specific context does not imply unobservability
in that very same context. For example, as is well known, the 
electrodynamic field can be altogether eliminated from the 
description of the dynamics of interacting charged 
point-particles~\cite{Wheeler.Feynman:1949}. But, surely, this is 
not saying that 
the field of one point particle cannot be measured by another one.} 
From that perspective one may indeed criticise special-relativistic scalar 
gravity for making essential use of dispensable absolute structures, 
which eventually should be eliminated, just like in the 
`flat-spacetime-approach' to GR; compare~\cite{Thirring:1961}
and Sect.\,5.2 in \cite{Giulini.Straumann:2006a}. 
In view of Quote\,\ref{quote:Einstein1} one might conjecture that 
this more sophisticated point was behind Einstein's criticism. 
If so, it is well taken. But physically it should be clearly 
separated from the other explicit accusations which we discussed 
here.   
   
\vspace{1.0truecm}
\addcontentsline{toc}{section}{Acknowledgements}
\noindent
\textbf{Acknowledgements:}
I thank two anonymous referees for making various suggestions 
for improvements and John Norton for asking a question that 
led to the remarks in the second part of Section\,\ref{sec:Conclusion}.
I am also indebted to Olivier Darrigol for pointing out that the 
argument leading to Remark\,\ref{rem:NoSimHitGround} in 
Section\,\ref{sec:Discussion} does not generalise to vector 
theories of gravity, as originally proposed in an earlier version 
of this paper; cf. footnote\,\ref{fnote:OliviersRemark}. 

\addcontentsline{toc}{section}{References}
\newpage

\end{document}